\documentclass[prd,aps,11pt,notitlepage,showpacs,nofootinbib,tightenlines]{revtex4}
\usepackage{mathrsfs}
\usepackage{amsmath}
\usepackage{amssymb}
\usepackage{epsfig}
\usepackage{graphicx}
\usepackage{booktabs}
\usepackage{multirow}
\usepackage{subfigure}
\usepackage{bm}
\usepackage{times}
\usepackage{braket}
\usepackage{color}
\usepackage{slashed}
\usepackage{hyperref}
\newcommand{\beq}{\begin{eqnarray}}
\newcommand{\eeq}{\end{eqnarray}}
\newcommand{\non}{\nonumber\\ }

\definecolor{Red}{rgb}{1.,0.,0.}

\definecolor{Blue}{rgb}{0.,0.,1.}
\newcommand{\Blue}[1]{{\color{Blue}{#1}}}

\definecolor{nicered}{rgb}{0.7,0.1,0.1}
\definecolor{nicegreen}{rgb}{0.1,0.5,0.1}
\hypersetup{colorlinks,citecolor=nicegreen,linkcolor=nicered}
\def \cpc{ Chin. Phys. C  }

\def \epjc{ Eur. Phys. J. C }

\def \jpg{  J. Phys. G }
\def \npb{  Nucl. Phys. B }
\def \plb{  Phys. Lett. B }
\def \ppnp{ Prog. Part. Nucl. Phys. }
\def \prd{  Phys. Rev. D }
\def \prl{  Phys. Rev. Lett.  }

\def \jhep{ J. High Energy Phys. }
\begin{document}

\title{Resonant contributions to three-body $B_{(s)} \to [ D^{(*)}, \bar{D}^{(*)} ] K^+K^-$ decays in the perturbative QCD approach}

\author{Ya Li$^1$~\footnote{correspondence author, liyakelly@163.com},
Da-Cheng Yan$^2$~\footnote{correspondence author, yandac@126.com},
Zhou Rui$^3$~\footnote{correspondence author, jindui1127@126.com},
Lan Liu$^{4}$,
Yue-Tong Zhang$^{4}$,
and Zhen-Jun Xiao$^{4}$~\footnote{correspondence author, xiaozhenjun@njnu.edu.cn}}

\affiliation{$^1$ Department of Physics, College of Sciences, Nanjing Agricultural University, Nanjing, Jiangsu 210095, P.R. China}
\affiliation{$^2$ School of Mathematics and Physics, Changzhou University, Changzhou, Jiangsu 213164, P.R. China}
\affiliation{$^3$ College of Sciences, North China University of Science and Technology,
                          Tangshan 063009,  P.R. China}
\affiliation{$^4$ Department of Physics and Institute of Theoretical Physics,
                          Nanjing Normal University, Nanjing, Jiangsu 210023, P.R. China}
\date{\today}

\begin{abstract}
In this work, we study the $S$, $P$ and $D$ wave resonance contributions to three-body decays $B_{(s)} \to [ D^{(*)}, \bar{D}^{(*)} ] K^+K^-$ by employing the perturbative QCD
 (PQCD) approach, where the kaon-kaon invariant mass spectra are dominated by the $f_0(980),f_0(1370),$ $\phi(1020),\phi(1680),f_2(1270),f^{\prime}_2(1525),f_2(1750)$ and $f_2(1950)$ resonances.
The $KK$ $S$-wave component $f_0(980)$ is modeled with the Flatt\'e formalism, while other resonances are described by the relativistic Breit-Wigner (BW)
line shape.
The corresponding decay channels are studied by constructing the kaon-kaon distribution amplitude $\Phi_{KK}$, which captures important final state interactions between the kaon pair in the resonant region.
We found that the PQCD predictions for the branching ratios for most considered decays
agree with currently available data within errors.
The associated polarization fractions of those vector-vector and vector-tensor decay modes are also predicted,
which are expected to be tested in the near future experiments.
The invariant mass spectra for the corresponding resonances in the $B_{(s)} \to [D^{(*)}, \bar{D}^{(*)} ] K^+K^-$ decays are well established,
which can be confronted with the precise data from the LHCb and Belle II experiments.
\end{abstract}

\pacs{13.25.Hw, 12.38.Bx, 14.40.Nd }
\maketitle

\section{Introduction}

Studies of nonleptonic $B$ meson decays are crucial for testing the standard model (SM), understanding the quantum chromodynamics (QCD)
and searching for the possible new physics beyond SM.
Testing the SM requires first to measure its free parameters precisely and to understand how to improve the precision of theoretical calculations.
Most of the free parameters of the SM are related to flavor, such as the Cabibbo-Kobayashi-Maskawa (CKM) angles $\alpha,\beta$ and $\gamma$.
The precise measurement of the angle $\gamma$ of the CKM unitarity triangle is a hot topic both in flavor physics theories and experiments.
An analysis of the decays $B_{(s)} \to \bar{D}^{(*)0}\phi$ opens possibilities to offer competitive experimental precision on the angle $\gamma$~\cite{plb253-483,prd69-113003,plb649-61}.
Although the charmed decays $B_{(s)}\to D^{(*)0}\phi$ are CKM suppressed compared with the $B_{(s)}\to\bar{D}^{(*)0}\phi$ decays, they are important in the
CKM angle $\gamma$ extraction method. Therefore, a much deeper understanding of the related phenomena is required.

On the experimental side, more and more detailed analysis on the three-body $B$ meson hadronic decays have been performed by the BABAR~
\cite{prd73-011103,prd76-051103,prl100-171803,prd85-054023,prd85-112010}, Belle~\cite{plb542-171,prd76-012006,prd91-032008}
and LHCb~\cite{prl109-131801,jhep02-043,plb727-403,jhep08-005,prd92-032002,jhep01-131,prd98-071103,prd98-072006} collaborations based on the large data sample.
The decay $B^0_s \to \bar{D}^0\phi$ was first observed by the LHCb collaboration~\cite{plb727-403}.
Meanwhile, a significant signal $B^+\to D^+_s K^+K^-$ is observed for the first time by the LHCb collaboration~\cite{jhep01-131}
and a limit of $\mathcal{B}(B^+\to D^+_s \phi)<4.9\times10^{-7}(4.2\times10^{-7})$ is set on the branching fraction at $95\% (90\%)$ confidence level (CL).
In addition, the LHCb collaboration~\cite{prd98-071103} reported their  first measurement ${\cal B }(B^0_s \to \bar{D}^{*0}\phi)=(3.7\pm 0.6)\times 10^{-5}$
and gave an upper limit ${\cal B}(B^0\to \bar{D}^0\phi) < 2.3 \times 10^{-6}$ at $95\%$ CL, where the
$\phi$ meson is reconstructed through its decay to a $K^+K^-$ pair.

On the theoretical side, the two-body charmed decays $B_{(s)}\to [ D^{(*)}, \bar{D}^{(*)}] [S, P, V, T]$ (here $S, P, V$ and $T$ denote the scalar, pseudoscalar, vector and tensor mesons) have been investigated within the framework of the PQCD factorization approach~\cite{prd67-054028,prd68-094018,HEPNP27-1062,prd78-014018,jpg37-015002,prd86-094001,epjc77-870}.
The channels induced by the $b\to c$ transitions are CKM favored, while those induced by the $b \to u$ transitions are CKM suppressed.
Thus, the $b \to u$ decays will have smaller branching ratios.
The interference between the $b\to c$ and $b \to u$ transitions gives the measurement of the CKM angle $\gamma$.
As is well-known, the charmed decays of $B_{(s)}$ are more complicated because of the hierarchy of the scale involved compared
with the decays of $B_{(s)}$ mesons to the light vector mesons.
For example, the $B\to D$ transitions involve three scales: the $B$ meson mass $m_B$, the $D$ meson mass $m_D$, and the mass difference from the heavy meson
and the heavy quark $\bar{\Lambda}=m_B-m_b\sim m_D-m_c$, which are strikingly different from each other.
Although, the factorization has been proved in soft-collinear effective theory~\cite{prd68-114009}, it needs more inputs than the PQCD approach.
It can be found that the momentum square of the hard gluon connecting the spectator quark is only a factor of $(1-m^2_D/m^2_B)$ to that of the $B \to$
light  transitions for $B\to D$ transitions, which ensures that PQCD can also work well in $B\to D$ transitions.
There are also many other traditional methods and approaches to estimate the $B \to D$ transitions, such as
the heavy quark effective theory (HQET)~\cite{prd95-115008,jhep12-060}, light cone sum rules (LCSR)~\cite{epjc60-603,jhep06-062,epjc78-76} and lattice QCD (LQCD)~\cite{1812.07675,epjw175-13012,1811-00794}.

As addressed before, the $B \to DKK$ decay is expected to proceed through $\phi \to KK$ intermediate state.
Moreover, this process can also be dominated by a series of other resonances in $S$, $P$, and $D$ waves.
In this work, we will study the $S$, $P$ and $D$ wave resonance contributions to three-body decays $B \to DKK$ by employing the PQCD approach,
where the kaon-kaon invariant mass spectra are dominated by the $f_0(980),f_0(1370),\phi(1020),\phi(1680),f_2(1270),f^{\prime}_2(1525), f_2(1750)$ and $f_2(1950)$ resonances.
However, the theory of three-body non-leptonic decays is still in an early stage of development.
Three-body $B$ meson decay modes do receive the entangled resonant and nonresonant contributions, as well as the possible final-state interactions
(FSIs)~\cite{prd89-094013,1512-09284,prd89-053015}, whereas the relative strength of these contributions vary significantly in different regions
of the Dalitz plots~\cite{dalitz-plot1,dalitz-plot2}.
In this respect, three-body decays are considerably more challenging than two-body decays, but provide a number of theoretical and
phenomenological advantages.
On the one hand, the number of different three-body final states is about ten times larger than the number of two-body decays.
On the other hand, each final state has a non-trivial kinematic multiplicity (a two-dimensional phase space) as opposed to two-body decays where the kinematics
is fixed by the masses.
This leads to a much richer phenomenology, but there is no proof of factorization for the three-body $B$ decays at present.
We can only restrict ourselves to specific kinematical configurations on basis of the Dalitz plot analysis.
The Dalitz plot contains different regions with ``specifical'' kinematics~\cite{npb899-247,1609-07430}.
The central region corresponds to the case where all three final particles fly apart with similar large energy and none of them moves collinearly to any other.
This situation contains two hard gluons and is power and $\alpha_s$ suppressed with respect to the amplitude at the edge.
The corners correspond to the case in which one final particle is approximately at rest (i.e. soft), and the other two are fast and  back-to-back.
The central part of the edges corresponds to the case in which two particles move collinearly and the other particle recoils back: this is called the quasi-two-body decay.
This situation exists particularly in the low $\pi\pi$ or $K\pi$ or $KK$ invariant mass region of the Dalitz plot.
Thereby, it is reasonable to assume the validity of factorization for the quasi-two-body $B$ meson decay.
Naturally the dynamics associated with the pair of final state mesons can be factorized into a two-meson distribution amplitude (DA)
$\Phi_{h_1h_2}$~\cite{MP,MT01,MT02,MT03,NPB555-231,Grozin01,Grozin02}.

In recent years, based on the PQCD approach, more and more detailed analysis on the three-body $B$ meson hadronic decays have
been performed in the low energy resonances on $\pi\pi$, $KK$ , $K\pi$ and $\pi\eta$ channels~
\cite{plb561-258,plb763-29,epjc76-675,cpc41-083105,epjc77-199,prd97-033006,prd95-056008,prd96-093011,epjc79-37,epjc79-539,jpg46-095001,epjc79-792,cpc44-073102,jhep03-162,2004-09027,epjc80-394,2005-02097,cpc43-073103,prd99-093007}.
Other theoretical approaches for describing the three-body  hadronic decays of $B$ mesons based on the symmetry principles
and factorization theorems have been developed.
The QCD-improved factorization (QCDF)~\cite{prl83-1914,npb591-313,npb606-245,npb675-333} has been widely adopted in the study
of the three-body charmless hadronic $B$ meson decays~\cite{npb899-247,plb622-207,prd74-114009,prd79-094005,APPB42-2013,prd76-094006,prd88-114014,prd94-094015,prd89-094007,prd87-076007,jhep10-117,2005-06080,prd99-076010}.
The $U$-spin and flavor $SU(3)$ symmetries were used in Refs.~\cite{prd72-094031,plb727-136,prd72-075013,prd84-056002,plb728-579,prd91-014029}.
Unlike the collinear factorization in the QCD factorization approach and soft-collinear
effective theory, the $k_T$ factorization is utilized in the PQCD approach.
In this approach, the transverse momentum of valence quarks in the mesons is kept to avoid the endpoint singularity~\cite{JHEP06-013,JHEP02-008}.
The Sudakov factors from the $k_T$ resummation have been included to suppress the long-distance contributions from the large-$b$ region
with $b$ being a variable conjugate to $k_T$. Therefore, one can calculate the color-suppressed channels
as well as the color-allowed channels in charmed $B$ decays within the PQCD approach.
The conventional non-calculable annihilation-type decays are also calculable in the PQCD approach, which is proved to be the
dominant strong phase in $B$ decays for the direct $CP$ asymmetry.

As aforementioned for the cases of the quasi-two-body decays, the two mesons ($h_1h_2$) move collinearly fast, and the
bachelor meson $h_3$ is also energetic and recoils against the meson pair in the $B$ meson rest frame in the quasi-two-body $B\to (h_1h_2)h_3$ decays.
The interaction between the meson pair and the bachelor meson is regarded as to be power suppressed.
The typical PQCD factorization formula for the $B\to (h_1h_2) h_3$ decay amplitude can be described as the form of ~\cite{plb561-258},
\begin{eqnarray}
\mathcal{A}=\Phi_B\otimes H\otimes \Phi_{h_1h_2}\otimes\Phi_{h_3},
\end{eqnarray}
where $H$ is the hard kernel, $\Phi_B$ and $\Phi_{h_3}$ are the universal wave functions of the $B$ meson and the bachelor
meson, respectively.
The hard kernel $H$ describes the dynamics of the strong and electroweak interactions in three-body hadronic decays in a similar way as the one for the corresponding two-body decays.
The wave functions $\Phi_B$ and $\Phi_{h_3}$ absorb the non-perturbative dynamics in the process.
The $\Phi_{h_1h_2}$ is the two-hadron ($KK$ pair in this work) DA, which involves the resonant and nonresonant interactions between the two moving collinearly mesons.

The present paper is organized as follows.
In Sec.~\ref{sec:2}, we give a brief introduction for the theoretical framework and
the total decay amplitudes with the wilson coefficients, CKM matrix elements and the amplitudes of four-quark operators needed in
the calculation will be given in Sec.~\ref{sec:3}.
Section~\ref{sec:4} contains the numerical values and some discussions.
A brief summary is given in section~\ref{sec:5}.
The Appendix collects the explicit PQCD factorization formulas for all the decay amplitudes.

\section{FRAMEWORK}\label{sec:2}
\subsection{The effective Hamiltonian and kinematics}\label{sec:21}
\begin{figure}[tbp]
\centerline{\epsfxsize=13cm \epsffile{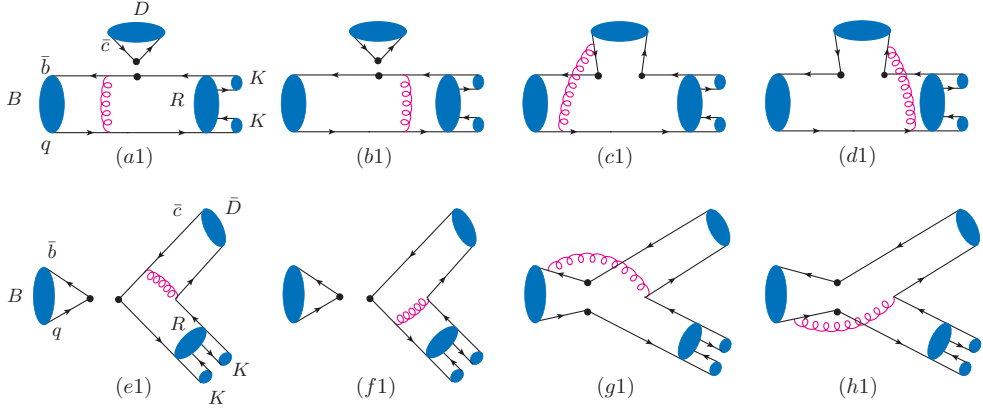}}
\caption{Typical leading-order Feynman diagrams for the quasi-two-body decays $B \to  \bar{D}^{(*)}(R\to) KK$,
with $q=(u,d,s)$, and the symbol $\bullet$ denotes the weak vertex.
 With the diagrams ($a$1)-($d$1) for the $B\to R\to KK$ transition, as well as the diagrams ($e$1)-($h$1) for annihilation contributions.}
\label{fig:fig1}
\end{figure}
\begin{figure}[tbp]
\centerline{\epsfxsize=13cm \epsffile{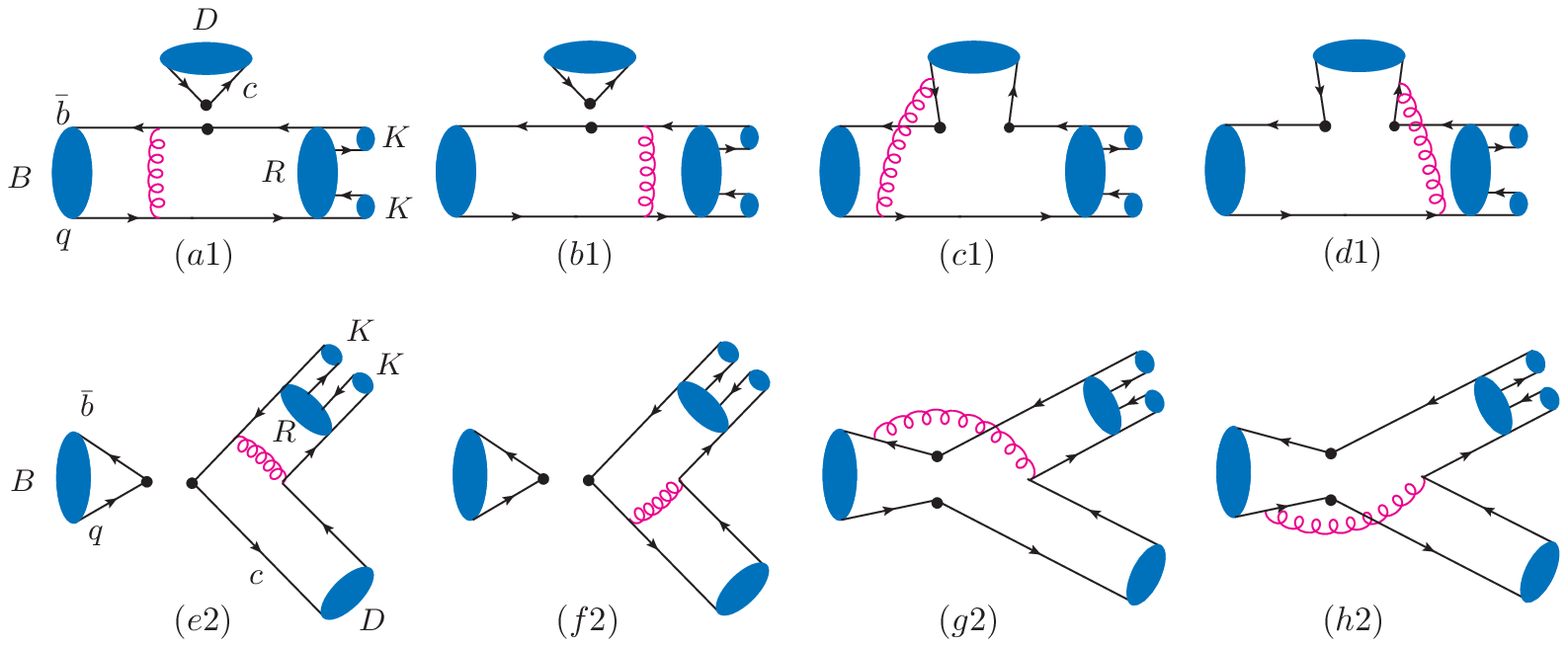}}
\caption{Typical leading-order Feynman diagrams for the quasi-two-body decays $B \to  D^{(*)}(R\to) KK$,
with $q=(u,d,s)$, and the symbol $\bullet$ denotes the weak vertex.
 With the diagrams ($a$1)-($d$1) for the $B\to R\to KK$ transition, as well as the diagrams ($e$2)-($h$2) for annihilation contributions.}
\label{fig:fig2}
\end{figure}
For $B \to D^{(*)}(R \to)KK$ decays, the weak effective Hamiltonian can be specified as follows~\cite{buras96}:
\begin{eqnarray}
{\cal  H}_{eff}&=& \left\{\begin{array}{ll}
\frac{G_F}{\sqrt{2}}V^*_{cb}V_{uq}\left[C_1(\mu)O_1(\mu)+C_2(\mu)O_2(\mu)\right],
& \ \  {\rm for} \ \ B_{(s)} \to  \bar{D}^{(*)}(R \to) KK \ \ {\rm decays},\\
\frac{G_F}{\sqrt{2}} V^*_{ub}V_{cq}\left[C_1(\mu)O_1(\mu)+C_2(\mu)O_2(\mu)\right],
& \ \  {\rm for} \ \ B_{(s)} \to  D^{(*)}(R \to) KK \ \ {\rm decays},\\
\end{array} \right.
 \label{eq:hamiltonian01}
\end{eqnarray}
where $V_{cb(q)}$ and $V_{ub(q)}$ are the CKM matrix elements and $R$ denotes the various partial wave resonances $f_0(980),f_0(1370),\phi(1020),\phi(1680),f_2(1270),f^{\prime}_2(1525),f_2(1750)$ and $f_2(1950)$~\footnote{ In the following, we also use the abbreviation $f_0$, $\phi$, and $f_2$
to denote the $S$, $P$, and $D$-wave resonances  for simplicity.}, respectively.
The explicit expressions of the local four-quark tree operators $O_{1,2}(\mu)$ and the corresponding Wilson coefficients $C_{1,2}(\mu)$ can be found in Ref.~\cite{buras96}.
The $q$ in Eq.~(\ref{eq:hamiltonian01}) represents the quark $d$ or $s$.
Noted that only tree diagrams contribute to these processes, which shows that there is no direct $CP$ asymmetry in these decays.
The typical Feynman diagrams for the decays $B_{(s)} \to  \bar{D}^{(*)}(R\to) KK$ and $B_{(s)} \to  D^{(*)}(R \to) KK$ are shown in Fig.~\ref{fig:fig1} and ~\ref{fig:fig2}, respectively.

We will work in the $B$ meson rest frame and employ the light-cone coordinates for momentum variables.
In the light-cone coordinates, we let the kaon pair and the final-state $D^{(*)}$ move along the directions $n=(1,0,0_{\rm T})$ and $v=(0,1,0_{\rm T})$, respectively.
The $B$ meson momentum $p_{B}$, the total momentum of the kaon pair, $p=p_1+p_2$, the final-state $D^{(*)}$ momentum $p_3$,  and the quark momentum $k_i$ in each meson are defined in the following form
\begin{eqnarray}
p_{B}&=&\frac{m_{B}}{\sqrt 2}(1,1,\textbf{0}_{\rm T}), ~\quad k_{B}=\left(0,x_B \frac{m_{B}}{\sqrt2} ,\textbf{k}_{B \rm T}\right),\nonumber\\
p&=&\frac{m_{B}}{\sqrt2}(1-r_D^2,\eta,\textbf{0}_{\rm T}), ~\quad k= \left( z (1-r_D^2)\frac{m_{B}}{\sqrt2},0,\textbf{k}_{\rm T}\right),\nonumber\\
p_3&=&\frac{m_{B}}{\sqrt 2}(r_D^2,1-\eta,\textbf{0}_{\rm T}), ~\quad k_3=\left(0,x_3(1-\eta) \frac{m_B}{\sqrt{2}},\textbf{k}_{3{\rm T}}\right),\label{mom-B-k}
\end{eqnarray}
where $m_{B}$ is the mass of the $B$ meson, $\eta=\frac{\omega^2}{m^2_{B}(1-r_D^2)}$ with $r_D=m_{D^{(*)}}/m_{B}$, $m_{D^{(*)}}$ is the mass of the bachelor meson, and the invariant mass squared $\omega^2=(p_1+p_2)^2=p^2$.
The momentum fractions $x_{B}$, $z$, and $x_3$ run from zero to unity, respectively.
In the heavy quark limit, the mass difference $\bar{\Lambda}$ between $b$-quark ($c$-quark) and $B(D)$ meson is negligible.

As usual we also define the momentum $p_1$ and $p_2$ of kaon pair as
\begin{eqnarray}\label{eq:p1p2}
 p_1&=&(\zeta p^+, (1-\zeta)\eta p^+, \textbf{p}_{1\rm T}), ~\quad p_2=((1-\zeta) p^+, \zeta\eta p^+, \textbf{p}_{2\rm T}),
\end{eqnarray}
with $\zeta=p_1^+/P^+$ characterizing the distribution of the longitudinal momentum of the kaon and $p_{1\rm T}^2=p_{2\rm T}^2=\zeta(1-\zeta)\omega^2$.

\subsection{Wave functions of $B$ meson and the $D^{(*)}$ mesons}\label{sec:22}

The light-cone matrix element of the $B$ meson  can be decomposed as~\cite{prd65-014007}:
\begin{eqnarray}
\int d^4z e^{i{\bf k_1}\cdot z}\langle0|q_{\beta}(z)\bar{b}_{\alpha}(0)|B(p_{B})\rangle=
\frac{i}{\sqrt{2N_c}}\left\{({p \hspace{-2.0truemm}/}_{B}+m_{B})\gamma_{5}\left[\phi_{B}({\bf k_1})-
\frac{{n \hspace{-2.0truemm}/}-{v \hspace{-2.0truemm}/}}{\sqrt{2}}\bar{\phi}_{B}({\bf k_1})\right]\right\}
_{\beta\alpha},
\end{eqnarray}
where $q$ represents $u$ or $d$ or $s$ quark.
According to the above equation, there are two different wave functions $\phi_{B}$ and $\bar \phi_{B}$ in the $B$ meson distribution amplitudes, which obey the following normalization conditions:
\begin{eqnarray}
\int \frac{d^4{\bf k_1}}{(2\pi)^4}\phi_{B}({\bf k_1})=\frac{f_{B}}{2\sqrt{6}},\;\;\int\frac{d^4{\bf k_1}}{(2\pi)^4}\bar{\phi}_{B}({\bf k_1})=0.
\end{eqnarray}

In general, one should consider these two Lorentz structures in the calculations of $B$ decays as shown in Refs.~\cite{prd85-074004}.
However, we neglect the contribution of $\bar \phi_{B}$ because of the numerical suppression in this work.
Then, the wave function of $B$ meson can be written as~\cite{prd63-054008,prd65-014007,epjc28-515,ppnp51-85,prd85-094003,plb504-6,prd63-074009}
\begin{eqnarray}
\Phi_B= \frac{i}{\sqrt{2N_c}} ({ p \hspace{-2.0truemm}/ }_B +m_B) \gamma_5 \phi_B ({\bf k_1}) \;,
\label{bmeson}
\end{eqnarray}
with the widely used $B$-meson DA in the PQCD approach~\cite{prd63-054008,ppnp51-85}
\begin{eqnarray}
\phi_B(x,b)&=& N_B x^2(1-x)^2\mathrm{exp} \left
 [ -\frac{M_B^2\ x^2}{2 \omega_{B}^2} -\frac{1}{2} (\omega_{B}\; b)^2\right] \;,
 \label{phib}
\end{eqnarray}
where the normalization factor $N_B$ depends on the values of $\omega_B$ and $f_B$ and defined through the normalization relation
$\int_0^1dx \; \phi_B(x,b=0)=f_B/(2\sqrt{6})$. $\omega_B$ is a free parameter and $\omega_B = 0.40 \pm
0.04$ GeV and $\omega_{B_s}=0.50 \pm 0.05$ GeV ~\cite{prd63-054008,plb504-6,prd63-074009} are used in the numerical calculations.
Very recently, a new method was proposed to calculate the $B$ meson light-cone DA from lattice QCD, which can be used as an updated input
for the $B$ meson DA in the future~\cite{prd102-011502}.

For the $D^{(*)}$ meson, in the heavy quark limit, the two-parton light-cone DA can be written as~\cite{prd67-054028,prd68-094018,HEPNP27-1062,prd78-014018,jpg37-015002,prd86-094001}
\begin{eqnarray}
\langle D(p_3)|q_{\alpha}(z)\bar{c}_{\beta}(0)|0\rangle
\,&=&\,\frac{i}{\sqrt{2N_{c}}}\int_{0}^{1}dx\,e^{ixp_3\cdot
z}\left[\gamma_{5}(\makebox[-1.5pt][l]{/}p\,_3+\,m_{D})\phi_{D}(x,b)\right]_{\alpha\beta},\\
\langle D^*(p_3)|q_{\alpha}(z)\bar{c}_{\beta}(0)|0\rangle
\,&=&\,-\frac{1}{\sqrt{2N_{c}}}\int_{0}^{1}dx\,e^{ixp_3\cdot
z}[{\epsilon\hspace{-1.5truemm}/}_L(\makebox[-1.5pt][l]{/}p\,_3+\,m_{D^*})\phi^L_{D^*}(x,b)\non
&+&{\epsilon\hspace{-1.5truemm}/}_T(\makebox[-1.5pt][l]{/}p\,_3+\,m_{D^*})\phi^T_{D^*}(x,b)]_{\alpha\beta},
\end{eqnarray}
 where
\begin{eqnarray}
\phi_{D}(x,b)=\phi_{D^*}^{L(T)}(x,b)=\frac{1}{2\sqrt{2N_{c}}}\,f_{D^{(*)}}\,6x(1-x)\left[1+C_{D}(1-2x)\right]
\exp\left[\frac{-\omega^{2}b^{2}}{2}\right],
\end{eqnarray}
with $C_{D}=0.5\pm0.1, \omega=0.1$ GeV and $f_{D^+}=211.9$ MeV~\cite{pdg2018} for $D$ meson.
In the above models, $x$ is the momentum fraction of the light quark in $D$ meson.
We determine the decay constant of the vector meson $D^*$ by using the relation $f_{D^*}=\sqrt{\frac{m_{D}}{m_{D^*}}}f_{D}$ based on the heavy quark effective theory~\cite{npc10-1}.

\subsection{Two-kaon DAs}

Below, we briefly introduce the $S$, $P$, and $D$-wave two-kaon DAs and the corresponding time-like form factors used in our framework.
It will be shown that resonant contributions through two-body channels can be included by parameterizing the two-kaon DAs.
The $S$-wave two-kaon DAs are described in the following form~\cite{epjc79-792},
\begin{eqnarray}\label{swave}
\Phi_{S}^{I=0}=\frac{1}{\sqrt{2N_c}}[{p\hspace{-2.0truemm}/}\phi^0_S(z,\zeta,\omega^2)+
\omega\phi^s_S(z,\zeta,\omega^2)+\omega({n\hspace{-2.0truemm}/}{v\hspace{-2.0truemm}/}-1)\phi^t_S(z,\zeta,\omega^2)].
\end{eqnarray}
In what follows the subscripts $S$, $P$, and $D$ are always associated with the corresponding partial waves.

Above various twists DAs have similar forms as the corresponding twists for a scalar meson
by replacing the scalar decay constant with the scalar form factor~\cite{plb730-336}.
For the scalar resonances $f_0(980)$ and $f_0(1370)$, the asymptotic forms of the individual DAs in Eq.~(\ref{swave}) have been parameterized as~\cite{MP,MT01,MT02,MT03}
\begin{eqnarray}
\phi^0_S(z,\zeta,\omega^2)&=&\frac{9F_S(\omega^2)}{\sqrt{2N_c}}a_Sz(1-z)(1-2z),\\
\phi^s_S(z,\zeta,\omega^2)&=&\frac{F_S(\omega^2)}{2\sqrt{2N_c}},\\
\phi^t_S(z,\zeta,\omega^2)&=&\frac{F_S(\omega^2)}{2\sqrt{2N_c}}(1-2z),
\end{eqnarray}
with the time-like scalar form factor $F_S(\omega^2)$ and the Gegenbauer coefficient $a_S$.

The $P$-wave two-pion DAs related to both longitudinal and transverse polarizations have been studied in Ref.~\cite{prd98-113003}.
Naively, the $P$-wave two-kaon ones can be obtained by replacing the pion vector form factors with the corresponding
kaon ones.
The explicit expressions of the $P$-wave kaon-kaon DAs associated with longitudinal (L) and transverse (T) polarization are described as follows,
\begin{eqnarray}
\Phi_P^{L}&=&\frac{1}{\sqrt{2N_c}} \left [{ p \hspace{-1.5truemm}/ }\phi_P^0(z,\zeta,\omega^2)+\omega\phi_P^s(z,\zeta,\omega^2)
+\frac{{p\hspace{-1.5truemm}/}_1{p\hspace{-1.5truemm}/}_2
  -{p\hspace{-1.5truemm}/}_2{p\hspace{-1.5truemm}/}_1}{\omega(2\zeta-1)}\phi_P^t(z,\zeta,\omega^2) \right ] \;,\label{pwavel}\\
\Phi_P^{T}&=&\frac{1}{\sqrt{2N_c}}
\Big [\gamma_5{\epsilon\hspace{-1.5truemm}/}_{T}{ p \hspace{-1.5truemm}/ } \phi_P^T(z,\zeta,\omega^2)
+\omega \gamma_5{\epsilon\hspace{-1.5truemm}/}_{T} \phi_P^a(z,\zeta,\omega^2)\non
&& + i\omega\frac{\epsilon^{\mu\nu\rho\sigma}\gamma_{\mu}
\epsilon_{T\nu}p_{\rho}n_{-\sigma}}{p\cdot n_-} \phi_P^v(z,\zeta,\omega^2) \Big ]\label{pwavet}.
\end{eqnarray}
The two-kaon DAs for various twists are expanded in terms of the Gegenbauer polynomials:
\begin{eqnarray}
\phi_P^0(z,\zeta,\omega^2)&=&\frac{3F_P^{\parallel}(\omega^2)}{\sqrt{2N_c}}z(1-z)\left[1
+a^0_{2P}\frac{3}{2}(5(1-2z)^2-1)\right] P_1(2\zeta-1) \;,\\
\phi_P^s(z,\zeta,\omega^2)&=&\frac{3F_P^{\perp}(\omega^2)}{2\sqrt{2N_c}}(1-2z)\left[1
+a^s_{2P}(10z^2-10z+1)\right] P_1(2\zeta-1) \;,\\
\phi_P^t(z,\zeta,\omega^2)&=&\frac{3F_P^{\perp}(\omega^2)}{2\sqrt{2N_c}}(1-2z)^2\left[1
+a^t_{2P}\frac{3}{2}(5(1-2z)^2-1)\right] P_1(2\zeta-1) \;,\\
\phi^T_P(z,\zeta,\omega^2)&=&\frac{3F_P^{\perp}(\omega^2)}
{\sqrt{2N_c}}z(1-z)\left[1+a^{T}_{2P}\frac{3}{2}(5(1-2z)^2-1)\right]\sqrt{\zeta(1-\zeta)},\\
\phi^a_P(z,\zeta,\omega^2)&=&\frac{3F_P^{\parallel}(\omega^2)}
{4\sqrt{2N_c}}(1-2z)\left[1+a^a_{2P}(10z^2-10z+1)\right]\sqrt{\zeta(1-\zeta)},\\
\phi^v_P(z,\zeta,\omega^2)&=&\frac{3F_P^{\parallel}(\omega^2)}
{8\sqrt{2N_c}}\bigg\{[1+(1-2z)^2]+a^v_{2P}[3(2z-1)^2-1]\bigg\}\sqrt{\zeta(1-\zeta)},
\end{eqnarray}
with the two $P$-wave form factors $F_P^{\parallel}(\omega^2)$ and $F_P^{\perp}(\omega^2)$ and the Gegenbauer coefficients $a^i_{2P}$.

We introduce the $D$-wave two-kaon DAs associated with longitudinal
and transverse polarizations as follows~\cite{epjc79-792},
\begin{eqnarray}\label{eq:dwavekpi}
\Phi_{D}^L&=&\sqrt{\frac{2}{3}}\frac{1}{\sqrt{2N_c}}
\left [ {p\hspace{-2.0truemm}/}\phi^0_D(z,\zeta,\omega^2)+\omega \phi^s_D(z,\zeta,\omega^2)+
\frac{{p\hspace{-1.5truemm}/}_1{p\hspace{-1.5truemm}/}_2-{p\hspace{-1.5truemm}/}_2{p\hspace{-1.5truemm}/}_1}{\omega(2\zeta-1)}
\phi^t_D(z,\zeta,\omega^2) \right ]\;,\label{dwavel}\\
\Phi_{D}^{T}&=&\sqrt{\frac{1}{2}}\frac{1}{\sqrt{2N_c}}
[\gamma_5{\epsilon\hspace{-1.5truemm}/}_{T}{p\hspace{-2.0truemm}/} \phi^T_D(z,\zeta,\omega^2)
+\omega \gamma_5{\epsilon\hspace{-1.5truemm}/}_{T} \phi^a_D(z,\zeta,\omega^2)\non
&&+i\omega\frac{\epsilon^{\mu\nu\rho\sigma}\gamma_{\mu}
\epsilon_{ T\nu}p_{\rho}n_{-\sigma}}{p\cdot n_-} \phi^v_D(z,\zeta,\omega^2)]\label{dwavet}\Blue{.}
\end{eqnarray}
The $D$-wave DAs are given as
\begin{eqnarray}\label{eq:dwavetwist3}
\phi^0_D(z,\zeta,\omega^2)&=&\frac{6F^{\parallel}_D(\omega^2)}{2\sqrt{2N_c}}z(1-z) \left [3a_{D}^0(2z-1) \right ] P_2(2\zeta-1),\label{eq:phid0}\\
\phi^s_D(z,\zeta,\omega^2)&=&-\frac{9F^{\perp}_D(\omega^2)}{4\sqrt{2N_c}}\left [ a_{D}^0(1-6z+6z^2) \right ] P_2(2\zeta-1),\label{eq:phids}\\
\phi^t_D(z,\zeta,\omega^2)&=&\frac{9F^{\perp}_D(\omega^2)}{4\sqrt{2N_c}}\left [ a_{D}^0(1-6z+6z^2)(2z-1) \right ] P_2(2\zeta-1),\label{eq:phidt}\\
\phi^T_D(z,\zeta,\omega^2)&=&\frac{6F^{\perp}_D(\omega^2)}{2\sqrt{2N_c}}z(1-z)\left[3a_D^T(2z-1)\right]\mathcal{T}(\zeta),\label{eq:tphid}\\
\phi^a_D(z,\zeta,\omega^2)&=&\frac{3F^{\parallel}_D(\omega^2)}{2\sqrt{2N_c}}a_D^T(2z-1)^3\mathcal{T}(\zeta),\label{eq:aphid}\\
\phi^v_D(z,\zeta,\omega^2)&=&-\frac{3F^{\parallel}_D(\omega^2)}{2\sqrt{2N_c}}a_D^T(1-6z+6z^2)\mathcal{T}(\zeta),\label{eq:vphid}
\end{eqnarray}
with the $\zeta$ dependent factor $P_2(2\zeta-1)=1-6\zeta(1-\zeta)$ and $\mathcal{T}(\zeta)=(2\zeta-1)\sqrt{\zeta(1-\zeta)}$.
$F^{\parallel,\perp}_D(\omega^2)$ are the $D$-wave time-like form factors and the Gegenbauer moments $a_{D}^{0,T}$ have been determined in our previous work~\cite{epjc79-792}.

The strong interactions between the resonance and the final-state meson pair, including elastic rescattering of the final-state meson pair, can be factorized into the time-like form factor $F_{S,P,D}(\omega^2)$, which is guaranteed by the Watson theorem~\cite{pr88-1163}.
For a narrow resonance, we usually use the relativistic BW line shape to parameterize the time-like form factor $F(\omega^2)$.
The explicit expression is~\cite{epjc78-1019},
\begin{eqnarray}
\label{BRW}
F(\omega^2)&=&\sum_i\frac{c_i m_i^2}{m^2_i -\omega^2-im_i\Gamma_i(\omega^2)} \;,
\end{eqnarray}
where the corresponding weight coefficients $c_i$ are determined based on the normalization condition $F(0)=1$.
The $m_i$ and $\Gamma_i$ are the pole mass and width of the corresponding resonances shown in Table~\ref{Tab:pa}, respectively.
The mass-dependent width $\Gamma_i(\omega)$ is defined as
\begin{eqnarray}
\label{BRWl}
\Gamma_i(\omega^2)&=&\Gamma_i\left(\frac{m_i}{\omega}\right)\left(\frac{|\vec{p}_1|}{|\vec{p}_0|}\right)^{(2L_R+1)}.
\end{eqnarray}
The $|\vec{p}_1|$ is the momentum vector of the resonance decay product measured in the resonance rest frame, while $|\vec{p}_0|$ is the value of $|\vec{p}_1|$ at $\omega=m_i$.
The explicit expression of kinematic variables $|\vec{p}_1|$ is
\begin{eqnarray}
|\vec{p}_1|=\frac{\sqrt{\lambda(\omega^2,m_{h_1}^2,m_{h_2}^2)}}{2\omega},
\end{eqnarray}
with the K$\ddot{a}$ll$\acute{e}$n function $\lambda (a,b,c)= a^2+b^2+c^2-2(ab+ac+bc)$.
$L_R$ is the orbital angular momentum in the dikaon system and $L_R=0,1,2,...$ corresponds to the $S,P,D,...$ partial-wave resonances.
Due to the limited studies on the form factor $F^{\perp}(\omega^2)$, we use the two decay
constants $f_i^{(T)}$ of the intermediate particle to estimate the form factor ratio $r^T=F^{\perp}(\omega^2)/F^{\parallel}(\omega^2)\approx (f_i^T/f_i)$.

The BW formula does not work well for $f_0(980)$, because its pole mass is close to the $K\bar{K}$ threshold.
The resulting line shape above and below the threshold of the intermediate particle is called Flatt\'e parametrization~\cite{plb63-228}.
If the coupling of a resonance to the channel opening nearby is very strong, the Flatt\'e parametrization shows a scaling invariance and does not allow for an extraction of individual partial decay widths.
If the scalar resonance lies under the $K\bar{K}$ threshold, the position of the peak in the mass spectrum does not coincide with the pole mass of the resonance.
To solve the problem, the finite width to the propagator of scalar resonance has been taken into consideration by Achasov's~\cite{prd56-203,prd70-111901,prd97-036015}, which is the one-loop contribution to the self-energy of the scalar resonance from the two-particle intermediate states.
More details can be found in Refs.~\cite{prd56-203,prd70-111901,prd97-036015,prd96-091501,prd56-4084,plb363-106}.
In addition, the exponential factor $F_{KK}=e^{-\alpha q^2_K}$ with $\alpha=(2.0\pm0.25) {\rm GeV}^{-2}$ is introduced above the $K\bar{K}$ threshold and serves to reduce the $\rho_{KK}$ factor as the invariant mass increases, where $q_K$ is the koan momentum in the kaon-kaon rest frame~\cite{prd78-074023}.
This parametrization decreases the $f_0(980)$ width above $K\bar{K}$ threshold slightly.
In this work, the invariant mass of the dikaon is above the $K^+K^-$ threshold, we have tested the impact of the self-energy correction and found it is small. 
Thus, we employ the modified Flatt\'e model suggested by Bugg~\cite{prd78-074023} following the LHCb collaboration~\cite{prd89-092006,prd90-012003},
\begin{eqnarray}
F(\omega^2)=\frac{m_{f_0(980)}^2}{m_{f_0(980)}^2-\omega^2-im_{f_0(980)}(g_{\pi\pi}\rho_{\pi\pi}+g_{KK}\rho_{KK}F^2_{KK})}\;.
\end{eqnarray}
The coupling constants $g_{\pi\pi}=0.167$ GeV and $g_{KK}=3.47g_{\pi\pi}$~\cite{prd89-092006,prd90-012003}
describe the $f_0(980)$ decay into the final states $\pi^+\pi^-$ and $K^+K^-$, respectively.
The phase space factors $\rho_{\pi\pi}$ and $\rho_{KK}$ read
as~\cite{prd87-052001,prd89-092006,plb63-228}
\begin{eqnarray}
\rho_{\pi\pi}=\frac23\sqrt{1-\frac{4m^2_{\pi^\pm}}{\omega^2}}
 +\frac13\sqrt{1-\frac{4m^2_{\pi^0}}{\omega^2}},\quad
\rho_{KK}=\frac12\sqrt{1-\frac{4m^2_{K^\pm}}{\omega^2}}
 +\frac12\sqrt{1-\frac{4m^2_{K^0}}{\omega^2}}.
\end{eqnarray}
\begin{table}  
\caption{Parameters used to describe intermediate states in our framework.}
\label{Tab:pa}
\begin{tabular*}{12cm}{@{\extracolsep{\fill}}llllll} \hline\hline
{\rm Resonance} &Mass~[MeV] &Width~[MeV] &$J^{PC}$&Model&Source\\ \hline
$f_0(980)$ &$990$ &$-$ &$0^{++}$&\rm Flatt\'e &PDG~\cite{pdg2018}\\
$f_0(1370)$ &1475&113 & $0^{++}$ & RBW &LHCb \cite{prd86-052006}\\
$\phi(1020)$ &1019&4.25& $1^{--}$ & RBW &PDG \cite{pdg2018}\\
$\phi(1680)$ &1689&211 & $1^{--}$ & RBW  &Belle \cite{prd80-031101}\\
$f_2(1270)$ &$1276$ &$187$ &$2^{++}$&\rm RBW &PDG~\cite{pdg2018}\\
$f'_2(1525)$ &1525&73 & $2^{++}$ & RBW &PDG \cite{pdg2018}\\
$f_2(1750)$ &1737&151 & $2^{++}$ & RBW &Belle \cite{epjc32-323}\\
$f_2(1950)$ &1980&297 & $2^{++}$ & RBW &Belle \cite{epjc32-323}\\
\hline\hline
\end{tabular*}
\end{table}
\subsection{The  differential branching ratio}
The double differential  branching ratio can be obtained as~\cite{pdg2018}
 \begin{eqnarray}\label{eq:br}
\frac{d^2\mathcal{B}}{d \zeta d\omega}=\frac{\tau_B\omega|\vec{p}_1||\vec{p}_3|}{32\pi^3 m_B^3}|\mathcal{A}|^2\Blue{.}
 \end{eqnarray}
The three-momenta of the kaon and $D$ meson in the $KK$ center-of-mass frame are given by
\begin{eqnarray}
|\vec{p}_1|=\frac{\sqrt{\lambda(\omega^2,m_K^2,m_{K}^2)}}{2\omega}, \quad
|\vec{p}_3|=\frac{\sqrt{\lambda(m_B^2,m_{D}^2,\omega^2)}}{2\omega}\;.
\end{eqnarray}
The complete amplitude $\mathcal{A}$ through intermediate resonances for the concerned decay channels can be written as the summation of
$\mathcal{A}_S$, $\mathcal{A}_P$ and $\mathcal{A}_D$:
 \begin{eqnarray}\label{eq:amplitude}
\mathcal{A}=\mathcal{A}_S+\mathcal{A}_P+\mathcal{A}_D,
 \end{eqnarray}
where $\mathcal{A}_S$, $\mathcal{A}_P$, and $\mathcal{A}_D$ denote the corresponding three $S$, $P$ and $D$ wave decay amplitudes.

Due to the angular momentum conservation requirement, the vector mesons $\phi,\bar D^*,D^*$ and tensor meson $f^{(\prime)}_2$ in the quasi-two-body decays
$B_{(s)} \to (\bar D^*, D^*)[f_0\to]KK$ and $B_{(s)} \to (\bar D, D)[\phi, f^{(\prime)}_2\to]KK$ should be completely polarized in the longitudinal direction.
For $B_{(s)}\to (\bar D^*, D^*)[\phi, f^{(\prime)}_2\to]KK$ decays, both the longitudinal polarization and the transverse polarization contribute.
The amplitudes can be decomposed as follows:
\begin{eqnarray}
\mathcal{A}_{P(D)}=\mathcal{A}_L+\mathcal{A}_N \epsilon_{T}\cdot \epsilon_{3T}
+i \mathcal{A}_T \epsilon_{\alpha\beta\rho\sigma} n_+^{\alpha} n_-^{\beta} \epsilon_{T}^{\rho} \epsilon_{3T}^{\sigma},
\end{eqnarray}
where $A_L$ is the longitudinally polarized decay amplitude, $A_N$ and $A_T$ are the transversely polarized contributions.
Therefore, the total decay amplitude for $B_{(s)}\to (\bar D^*, D^*)[\phi, f^{(\prime)}_2\to]KK$ decays can be expressed as
\begin{eqnarray}
|\mathcal{A}_{P(D)}|^2=|\mathcal{A}_{0}|^2+|\mathcal{A}_{\parallel}|^2+|\mathcal{A}_{\perp}|^2,
\end{eqnarray}
where $\mathcal{A}_0,\mathcal{A}_{\parallel}$ and $\mathcal{A}_{\perp}$ are defined as:
\begin{eqnarray}
\mathcal{A}_0=\mathcal{A}_L, \quad \mathcal{A}_{\parallel}=\sqrt{2}\mathcal{A}_{N},
\quad \mathcal{A}_{\perp}=\sqrt{2}\mathcal{A}_{T}.
\end{eqnarray}
The polarization fractions $f_{\lambda}$ with $\lambda=0$, $\parallel$, and $\perp$ are described as
\begin{eqnarray}\label{pol}
f_{\lambda}=\frac{|\mathcal{A}_{\lambda}|^2}{|\mathcal{A}_0|^2
+|\mathcal{A}_{\parallel}|^2+|\mathcal{A}_{\perp}|^2},
\end{eqnarray}
with the normalisation relation $f_0+f_{\parallel}+f_{\perp}=1$.
\section{Calculation of decay amplitudes in PQCD approach}\label{sec:3}
In this section, we intend to calculate the relevant decay amplitudes.
For the considered $B_{(s)}\to \bar D^{(*)}(R\to)KK$ decays, the analytic formulas for the corresponding decay amplitudes are of the
following form:
\begin{enumerate}
\item[$\bullet$] S-wave:
\begin{eqnarray}
{\cal A}(B_s^0\to \bar D^0(f_0\to)K^+K^-)&=&\frac{G_F}{\sqrt{2}}V^*_{cb}V_{us} \left [ a_2F^{LL}_{ef_0}+C_2M^{LL}_{ef_0} \right ],\\
{\cal A}(B_s^0\to \bar D^{*0}(f_0\to)K^+K^-)&=&\frac{G_F}{\sqrt{2}}V^*_{cb}V_{us}\left [ a_2F^{ LL}_{ef_0}+C_2M^{ LL}_{ef_0} \right].
\end{eqnarray}

\item[$\bullet$] P-wave:
\begin{eqnarray}
{\cal A}(B_s^0\to \bar D^0(\phi\to)K^+K^-)&=&\frac{G_F}{\sqrt{2}}V^*_{cb}V_{us}\left [a_2F^{LL}_{e\phi}+C_2M^{LL}_{e\phi} \right ],
\\
{\cal A}_i(B_s^0\to \bar D^{*0}(\phi\to)K^+K^-)&=&\frac{G_F}{\sqrt{2}}V^*_{cb}V_{us}\left [a_2F^{LL}_{e\phi,i}+C_2M^{LL}_{e\phi,i} \right ].
\end{eqnarray}

\item[$\bullet$] D-wave:
\begin{eqnarray}
{\cal A}(B^0\to \bar D^0(f^q_2\to)K^+K^-)&=&\frac{G_F}{\sqrt{2}}V^*_{cb}V_{ud}\Big [a_2(F^{LL}_{ef_2}+F^{LL}_{af_2})
+C_2(M^{LL}_{ef_2}+M^{LL}_{af_2}) \Big ],\\
{\cal A}(B_s^0\to \bar D^0(f^q_2\to)K^+K^-)&=&\frac{G_F}{\sqrt{2}}V^*_{cb}V_{us} \left [a_2F^{LL}_{af_2}+C_2M^{LL}_{af_2} \right ],\\
{\cal A}(B_s^0\to \bar D^0(f^s_2\to)K^+K^-)&=&\frac{G_F}{\sqrt{2}}V^*_{cb}V_{us} \left [a_2F^{LL}_{ef_2}+C_2M^{LL}_{ef_2} \right ],\\
{\cal A}_i(B^0\to \bar D^{*0}(f^q_2\to)K^+K^-)&=&\frac{G_F}{\sqrt{2}}V^*_{cb}V_{ud}\Big [a_2(F^{LL}_{ef_2,i}+F^{LL}_{af_2,i})
                                           +C_2(M^{LL,i}_{ef_2,i}+M^{LL,i}_{af_2,i}) \Big],\non \\
{\cal A}_i(B_s^0\to \bar D^{*0}(f^q_2\to)K^+K^-)&=&\frac{G_F}{\sqrt{2}}V^*_{cb}V_{us} \left [a_2F^{LLi}_{af_2,i}+C_2M^{LL}_{af_2,i} \right ],\\
{\cal A}_i(B_s^0\to \bar D^{*0}(f^s_2\to)K^+K^-)&=&\frac{G_F}{\sqrt{2}}V^*_{cb}V_{us} \left [a_2F^{LL}_{ef_2,i}+C_2M^{LL}_{ef_2,i} \right ].
\end{eqnarray}
\end{enumerate}

While for $B_{(s)}\to D^{(*)}(R\to)KK$ channels, the total decay amplitudes are written as:
\begin{enumerate}
\item[$\bullet$] S-wave:
\begin{eqnarray}
{\cal A}(B_s^0\to D^0(f_0\to)K^+K^-)&=&\frac{G_F}{\sqrt{2}}V^*_{ub}V_{cs}[a_2F^{LL}_{ef_0}+C_2M^{LL}_{ef_0}],
\\
{\cal A}(B_s^0\to D^{*0}(f_0\to)K^+K^-)&=&\frac{G_F}{\sqrt{2}}V^*_{ub}V_{cs}[a_2F^{LL }_{ef_0}+C_2M^{ LL}_{ef_0}].
\end{eqnarray}

\item[$\bullet$] P-wave:
\begin{eqnarray}
{\cal A}(B_s^0\to D^0(\phi\to)K^+K^-)&=&\frac{G_F}{\sqrt{2}}V^*_{ub}V_{cs}[a_2F^{LL}_{e\phi}+C_2M^{LL}_{e\phi}],
\\
{\cal A}_i(B_s^0\to D^{*0}(\phi\to)K^+K^-)&=&\frac{G_F}{\sqrt{2}}V^*_{ub}V_{cs}[a_2F^{LL}_{e\phi,i}+C_2M^{LL}_{e\phi,i}].
\end{eqnarray}

\item[$\bullet$] D-wave:
\begin{eqnarray}
{\cal A}(B^+\to D^+(f^q_2\to)K^+K^-)&=&\frac{G_F}{\sqrt{2}}V^*_{ub}V_{cd}[a_1(F^{LL}_{ef_2}+F^{LL}_{aD})
               +C_1(M^{LL}_{ef_2}+M^{LL}_{aD})],\\
{\cal A}(B^0\to D^0(f^q_2\to)K^+K^-)&=&\frac{G_F}{\sqrt{2}}V^*_{ub}V_{cd}[a_2(F^{LL}_{ef_2}+F^{LL}_{aD})
                                       +C_2(M^{LL}_{ef_2}+M^{LL}_{aD})], \quad \\
{\cal A}(B_s^0\to D^0(f^q_2\to)K^+K^-)&=&\frac{G_F}{\sqrt{2}}V^*_{ub}V_{cs}[a_2F^{LL}_{aD}+C_2M^{LL}_{aD}],\\
{\cal A}(B_s^0\to D^0(f^s_2\to)K^+K^-)&=&\frac{G_F}{\sqrt{2}}V^*_{ub}V_{cs}[a_2F^{LL}_{ef_2}+C_2M^{LL}_{ef_2}],\\
{\cal A}_i(B^+\to D^{*+}(f^q_2\to)K^+K^-)&=&\frac{G_F}{\sqrt{2}}V^*_{ub}V_{cd}[a_1(F^{LL}_{ef_2,i}+F^{LL}_{aD^*,i})
                                       +C_1(M^{LL}_{ef_2,i}+M^{LL}_{aD^*,i})], \quad \\
{\cal A}_i(B^0\to D^{*0}(f^q_2\to)K^+K^-)&=&\frac{G_F}{\sqrt{2}}V^*_{ub}V_{cd}[a_2(F^{LL}_{ef_2,i}+F^{LL}_{aD^*,i})
                                       +C_2(M^{LL}_{ef_2,i}+M^{LL}_{aD^*,i})],\quad \\
{\cal A}_i(B_s^0\to D^{*0}(f^q_2\to)K^+K^-)&=&\frac{G_F}{\sqrt{2}}V^*_{ub}V_{cs}[a_2F^{LL}_{aD^*,i}+C_2M^{LL}_{aD^*,i}],\\
{\cal A}_i(B_s^0\to D^{*0}(f^s_2\to)K^+K^-)&=&\frac{G_F}{\sqrt{2}}V^*_{ub}V_{cs}[a_2F^{LL}_{ef_2,i}+C_2M^{LL}_{ef_2,i}],
\end{eqnarray}
\end{enumerate}
in which the combinations $a_i$ of the Wilson coefficients are defined as:
\begin{eqnarray}
a_1=C_2+\frac{C_1}{3},\quad \quad \quad \quad \quad a_2=C_1+\frac{C_2}{3}.
\end{eqnarray}
For $B_{(s)}\to (\bar D^*, D^*)[\phi, f^{(\prime)}_2\to]KK$ decays, the superscript $i=L,N,T$, represent longitudinal, parallel, and transverse polarization contributions respectively.
$F^{LL}_{e(a)}$ and $ M^{LL}_{e(a)}$ describe the contributions from the factorizable emission (annihilation) and non-factorizable
emission (annihilation) diagrams as show in Fig.~\ref{fig:fig1} and~\ref{fig:fig2}.
The explicit expressions of the amplitudes $F^{LL}_{e(a)}$ and $ M^{LL}_{e(a)}$ will be given in the Appendix.

\section{Numerical results}\label{sec:4}

In this section, let us firstly list the parameters used in our numerical calculations, such as the masses (in units of GeV)~\cite{pdg2018}:
\begin{eqnarray}
m_{B}&=&5.280, \quad m_{B_s}=5.367, \quad m_b=4.8, \quad m_c=1.275,\quad m_{K^\pm}=0.494, \nonumber\\
m_{D^\pm}&=&1.870,  \quad m_{D^0/\bar{D}^0}=1.864,\quad m_{D^{*\pm}}=2.010,\quad  m_{D^{*0}/\bar{D}^{*0}}=2.007.
\end{eqnarray}
The values of the Wolfenstein parameters are adopted as given in the Ref.~\cite{pdg2018}:
$A=0.836\pm0.015, \lambda=0.22453\pm 0.00044$, $\bar{\rho} = 0.122^{+0.018}_{-0.017}$, $\bar{\eta}= 0.355^{+0.012}_{-0.011}$.

The decay constants (in units of GeV) and the $B$ meson lifetimes (in units of ps) are chosen
as~\cite{prd95-056008,epjc77-610,prd82-054019}
\begin{eqnarray}
f_B&=&0.19\pm0.02, \quad f_{B_s}=0.23\pm0.02, \quad f_{\phi(1020)}=0.215, \quad f_{\phi(1020)}^T=0.186,\nonumber\\
f_{f_2(1270)}&=&0.102, \quad f_{f_2(1270)}^T=0.117,\quad f_{f'_2(1525)}=0.126, \quad f_{f'_2(1525)}^T=0.065, \nonumber\\
\tau_{B^0}&=&1.519,\quad \tau_{B^{\pm}}=1.638, \quad \tau_{B_{s}}=1.512.
\end{eqnarray}

The form factor ratio $r^T$, the coefficients $c_i$ and the Gegenbauer moments $a_{S,P,D}$ are adopted the same values as those determined
in Ref.~\cite{epjc79-792}
\begin{eqnarray}\label{eq:rt1}
r^T(\phi(1020))&=&0.865, \quad r^T(\phi(1680))=0.6, \quad r^T(f_2(1270))=1.15, \nonumber\\
r^T(f'_2(1525))&=&0.52, \quad r^T(f_2(1750))=0.3, \quad r^T(f_2(1950))=1.5,\nonumber\\
c_{f_0(1370)}&=&0.12e^{i\frac{\pi}{2}}, \quad c_{\phi(1680)}=0.6, \quad c_{f'_2(1525)}=1.2, \nonumber\\
c_{f_2(1270)}&=&0.1e^{i\pi},  \quad c_{f_2(1750)}=0.4e^{i\pi},  \quad c_{f_2(1950)}=0.3, \nonumber\\
a_S&=&0.80\pm 0.16, \quad a_D^0=0.40\pm 0.08, \quad a_D^T=0.90\pm0.18\Blue{,} \nonumber\\
a^0_{2P}&=&-0.50\pm0.10, \quad a^s_{2P}=-0.70\pm0.14,\quad a^t_{2P}=-0.30\pm0.06,\nonumber\\
a^{T}_{2P}&=&-0.50\pm0.10, \quad a^a_{2P}=0.40\pm0.08,\quad a^v_{2P}=-0.60\pm0.12.
\end{eqnarray}

\begin{table}[!!b]  
\caption{PQCD results for the branching ratios of the $S$, $P$ and $D$ wave resonance channels in the $B^0_{(s)} \to \bar {D} K^+K^-$ decay together with experimental data~\cite{pdg2018}.
The theoretical errors are attributed to the variation of the shape parameters $\omega_{B_{(s)}}(C_{\bar{D}})$ in the wave function of $B_{(s)}(\bar{D})$ meson and the decay constant $f_{B_{(s)}}$,
the Gegenbauer moments of two-kaon DAs, and the hard scale $t$ and the QCD scale $\Lambda_{\rm QCD}$, respectively.}
\label{Tab:Db}
\begin{tabular*}{14cm}{@{\extracolsep{\fill}}llll} \hline\hline
{\rm Decay Modes} & & Quasi-two-body& Data~\cite{pdg2018}\\ \hline
$B_s^0\to  \bar D^{0}(f_0(980)\to)K^+K^-$&${\cal B}(10^{-6})$ &$1.36^{+0.61+0.67+0.79}_{-0.46-0.54-0.64}$&$<1.55$   \\
$B_s^0\to  \bar D^{0}(f_0(1370)\to)K^+K^-$&${\cal B}(10^{-7})$ &$5.66^{+2.68+2.70+3.18}_{-1.95-2.21-2.56}$&$\cdots$   \\
$B^0_s \to \bar{D}^0(\phi(1020)\to)K^+K^-$ &${\cal B}(10^{-5})$&$1.21^{+0.51+0.23+0.90}_{-0.41-0.30-0.54}$&$1.50\pm0.25$\\
$B^0_s \to \bar{D}^0(\phi(1680)\to)K^+K^-$ &${\cal B}(10^{-6})$&$1.35^{+0.67+0.24+1.10}_{-0.48-0.23-0.72}$&$\cdots$\\
$B^0_{s} \to \bar{D}^0 (f_{2}(1270)\to)K^+K^- $&${\cal B}(10^{-7})$&   $1.87^{+0.49+0.79+0.35}_{-0.47-0.69-0.62}$&$\cdots$    \\
$B^0_{s} \to \bar{D}^0 (f^{\prime}_{2}(1525)\to)K^+K^- $&${\cal B}(10^{-6})$&   $3.56^{+1.42+1.57+1.94}_{-1.12-1.28-1.62}$&$\cdots$     \\
$B^0_{s} \to \bar{D}^0 (f_{2}(1750)\to)K^+K^- $&${\cal B}(10^{-7})$&   $2.76^{+1.17+1.22+1.46}_{-0.90-0.99-1.18}$&$\cdots$     \\
$B^0_{s} \to \bar{D}^0 (f_{2}(1950)\to)K^+K^-$&${\cal B}(10^{-8})$&   $7.02^{+2.36+3.10+3.66}_{-2.01-2.52-3.91}$&$\cdots$    \\
$B^0 \to \bar{D}^0 (f_{2}(1270)\to)K^+K^- $&${\cal B}(10^{-6})$&   $3.54^{+1.60+1.55+1.20}_{-1.45-1.28-1.15}$&$\cdots$    \\
$B^0 \to \bar{D}^0 (f^{\prime}_{2}(1525)\to)K^+K^- $&${\cal B}(10^{-8})$&   $5.86^{+5.57+3.97+2.68}_{-1.53-1.49-1.90}$&$\cdots$    \\
\hline\hline
\end{tabular*}
\end{table}
\begin{table}[thb]  
\caption{PQCD results for the branching ratios of the $S$, $P$ and $D$ wave resonance channels in the  $B^0_{(s)} \to \bar {D}^* K^+K^-$ decay together with experimental data~\cite{pdg2018}.
The theoretical errors are attributed to the variation of the shape parameters $\omega_{B_{(s)}}(C_{\bar {D}^*})$ in the wave function of $B_{(s)}(\bar {D}^*)$ meson and the decay
constant $f_{B_{(s)}}$,
the Gegenbauer moments of two-kaon DAs, and the hard scale $t$ and the QCD scale $\Lambda_{\rm QCD}$, respectively.}
\label{Tab:Dsb}
\begin{tabular*}{14cm}{@{\extracolsep{\fill}}llll} \hline\hline
{\rm Decay Modes} & & Quasi-two-body& Data~\cite{pdg2018}\\ \hline
$B_s^0\to  \bar D^{*0}(f_0(980)\to)K^+K^-$&${\cal B}(10^{-6})$ &$1.14^{+0.52+0.56+0.66}_{-0.37-0.44-0.50}$&$\cdots$   \\
$B_s^0\to  \bar D^{*0}(f_0(1370)\to)K^+K^-$&${\cal B}(10^{-7})$ &$4.66^{+2.29+2.23+2.78}_{-1.55-1.82-2.09}$&$\cdots$\\
$B^0_s \to \bar{D}^{*0}(\phi(1020)\to)K^+K^-$ &${\cal B}(10^{-5})$&$1.76^{+0.64+0.22+0.92}_{-0.51-0.18-0.70}$&$1.85\pm0.30$\\
$B^0_s \to \bar{D}^{*0}(\phi(1680)\to)K^+K^-$ &${\cal B}(10^{-6})$&$2.11^{+0.90+0.27+0.98}_{-0.66-0.19-0.66}$&$\cdots$\\
$B^0_{s} \to \bar{D}^{*0} (f_{2}(1270)\to)K^+K^- $&${\cal B}(10^{-7})$&   $2.45^{+0.67+0.95+0.25}_{-0.56-0.78-0.31}$&$\cdots$     \\
$B^0_{s} \to \bar{D}^{*0} (f^{\prime}_{2}(1525)\to)K^+K^- $&${\cal B}(10^{-6})$&   $6.99^{+2.91+2.24+1.77}_{-2.19-1.81-1.35}$&$\cdots$     \\
$B^0_{s} \to \bar{D}^{*0} (f_{2}(1750)\to)K^+K^- $&${\cal B}(10^{-7})$&   $6.25^{+2.54+2.04+1.21}_{-1.98-1.66-0.99}$&$\cdots$     \\
$B^0_{s} \to \bar{D}^{*0} (f_{2}(1950)\to)K^+K^- $&${\cal B}(10^{-7})$&   $3.55^{+1.22+1.26+1.10}_{-1.05-1.03-0.74}$&$\cdots$     \\
$B^0 \to \bar{D}^{*0} (f_{2}(1270)\to)K^+K^- $&${\cal B}(10^{-5})$&   $3.67^{+1.21+1.42+0.66}_{-1.01-1.15-0.58}$&$\cdots$     \\
$B^0 \to \bar{D}^{*0} (f^{\prime}_{2}(1525)\to)K^+K^-$&${\cal B}(10^{-7})$&   $5.76^{+1.81+2.17+0.84}_{-1.57-1.77-0.59}$&$\cdots$    \\
\hline\hline
\end{tabular*}
\end{table}
\begin{table}  
\caption{PQCD results for the branching ratios of the $S$, $P$ and $D$ wave resonance channels in the  $B^0_{(s)} \to D K^+K^-$ decay.
The theoretical errors are attributed to the variation of the shape parameters $\omega_{B_{(s)}}(C_{D})$ in the wave function of $B_{(s)}(D)$ meson and the decay constant $f_{B_{(s)}}$,
the Gegenbauer moments of two-kaon DAs, and the hard scale $t$ and the QCD scale $\Lambda_{\rm QCD}$, respectively.}
\label{Tab:D}
\begin{tabular*}{12cm}{@{\extracolsep{\fill}}lll} \hline\hline
{\rm Decay Modes} & & Quasi-two-body\\ \hline
$B_s^0\to  D^{0}(f_0(980)\to)K^+K^-$&${\cal B}(10^{-7})$ &$1.30^{+0.62+0.58+0.31}_{-0.46-0.46-0.45}$ \\
$B_s^0\to  D^{0}(f_0(1370)\to)K^+K^-$&${\cal B}(10^{-8})$ &$3.91^{+1.54+1.89+0.48}_{-1.26-1.46-1.10}$   \\
$B^0_s \to D^0(\phi(1020)\to)K^+K^-$ &${\cal B}(10^{-7})$ &$9.65^{+3.64+4.05+2.03}_{-3.69-4.23-3.88}$\\
$B^0_s \to D^0(\phi(1680)\to)K^+K^-$ &${\cal B}(10^{-8})$&$7.42^{+4.32+4.27+2.49}_{-2.79-3.18-2.61}$\\
$B^0_{s} \to D^0 (f_{2}(1270)\to)K^+K^- $&${\cal B}(10^{-8})$&   $7.23^{+1.90+3.19+1.09}_{-1.58-2.60-1.37}$    \\
$B^0_{s} \to D^0 (f^{\prime}_{2}(1525)\to)K^+K^- $&${\cal B}(10^{-7})$&   $4.39^{+2.04+1.94+1.17}_{-1.53-1.58-1.54}$     \\
$B^0_{s} \to D^0 (f_{2}(1750)\to)K^+K^- $&${\cal B}(10^{-8})$&   $2.29^{+2.12+1.74+1.47}_{-0.39-0.50-0.49}$    \\
$B^0_{s} \to D^0 (f_{2}(1950)\to)K^+K^- $&${\cal B}(10^{-8})$&   $4.07^{+1.63+1.79+0.60}_{-1.28-1.47-1.20}$   \\
$B^+ \to D^+ (f_{2}(1270)\to)K^+K^-$&${\cal B}(10^{-7})$&   $1.76^{+0.44+0.78+0.10}_{-0.40-0.63-0.22}$   \\
$B^+ \to D^+ (f^{\prime}_{2}(1525)\to)K^+K^- $&${\cal B}(10^{-9})$&   $1.05^{+0.25+0.46+0.02}_{-0.24-0.38-0.15}$  \\
$B^0 \to D^0 (f_{2}(1270)\to)K^+K^- $&${\cal B}(10^{-8})$&   $1.66^{+0.56+0.73+0.16}_{-0.45-0.60-0.28}$   \\
$B^0 \to D^0 (f^{\prime}_{2}(1525)\to)K^+K^- $&${\cal B}(10^{-10})$&   $1.52^{+0.48+0.67+0.16}_{-0.37-0.54-0.29}$    \\
\hline\hline
\end{tabular*}
\end{table}

\begin{table}[thb]  
\caption{PQCD results for the branching ratios of the $S$, $P$ and $D$ wave resonance channels in the  $B^0_{(s)} \to D^* K^+K^-$ decay.
The theoretical errors are attributed to the variation of the shape parameters $\omega_{B_{(s)}}(C_{D^*})$ in the wave function of $B_{(s)}(D^*)$ meson and the decay constant $f_{B_{(s)}}$,
the Gegenbauer moments of two-kaon DAs, and the hard scale $t$ and the QCD scale $\Lambda_{\rm QCD}$, respectively.}
\label{Tab:Ds}
\begin{tabular*}{12cm}{@{\extracolsep{\fill}}lll} \hline\hline
{\rm Decay Modes} & & Quasi-two-body\\ \hline
$B_s^0\to  D^{*0}(f_0(980)\to)K^+K^-$&${\cal B}(10^{-8})$ &$9.67^{+4.23+3.21+2.23}_{-3.40-2.81-3.17}$  \\
$B_s^0\to  D^{*0}(f_0(1370)\to)K^+K^-$&${\cal B}(10^{-8})$ &$3.06^{+1.13+1.13+0.46}_{-1.00-0.97-0.81}$\\
$B^0_s \to D^{*0}(\phi(1020)\to)K^+K^-$ &${\cal B}(10^{-7})$&$6.39^{+1.73+0.93+0.07}_{-2.90-2.64-3.46}$\\
$B^0_s \to D^{*0}(\phi(1680)\to)K^+K^-$ &${\cal B}(10^{-8})$&$4.69^{+2.83+2.01+1.18}_{-1.66-1.82-1.54}$\\
$B^0_{s} \to D^{*0} (f_{2}(1270)\to)K^+K^- $&${\cal B}(10^{-8})$&   $3.12^{+0.67+1.18+0.62}_{-0.64-0.96-0.89}$    \\
$B^0_{s} \to D^{*0} (f^{\prime}_{2}(1525)\to)K^+K^- $&${\cal B}(10^{-7})$&   $4.01^{+1.61+1.64+0.79}_{-1.43-1.34-1.13}$     \\
$B^0_{s} \to D^{*0} (f_{2}(1750)\to)K^+K^- $&${\cal B}(10^{-8})$&   $2.22^{+1.11+0.93+0.65}_{-0.80-0.77-0.66}$   \\
$B^0_{s} \to D^{*0} (f_{2}(1950)\to)K^+K^- $&${\cal B}(10^{-8})$&   $4.87^{+1.93+1.81+0.61}_{-1.49-1.49-1.09}$    \\
$B^+ \to D^{*+} (f_{2}(1270)\to)K^+K^- $&${\cal B}(10^{-7})$&   $1.72^{+0.47+0.62+0.18}_{-0.42-0.51-0.20}$   \\
$B^+ \to D^{*+} (f^{\prime}_{2}(1525)\to)K^+K^- $&${\cal B}(10^{-9})$&   $1.06^{+0.24+0.36+0.10}_{-0.24-0.30-0.13}$    \\
$B^0 \to D^{*0} (f_{2}(1270)\to)K^+K^- $&${\cal B}(10^{-8})$&   $1.10^{+0.36+0.38+0.17}_{-0.30-0.32-0.30}$     \\
$B^0 \to D^{*0} (f^{\prime}_{2}(1525)\to)K^+K^- $&${\cal B}(10^{-11})$&   $7.96^{+2.41+3.21+1.12}_{-2.11-2.61-2.61}$    \\
\hline\hline
\end{tabular*}
\end{table}
\begin{table} [thb] 
\caption{PQCD results for the polarization fractions of the $B^0_{(s)} \to (\bar D^{*}, D^{*})[\phi, f^{(\prime)}_2\to] K^+K^-$  decays together with experimental data~\cite{pdg2018}.
The theoretical errors are attributed to the variation of the shape parameters $\omega_{B_{(s)}}(C_{D^*})$ in the wave function of $B_{(s)}(D^*)$ meson and the decay
constant $f_{B_{(s)}}$,
the Gegenbauer moments of two-kaon DAs, and the hard scale $t$ and the QCD scale $\Lambda_{\rm QCD}$, respectively.}
\label{Tab:pola}
\begin{tabular*}{15cm}{@{\extracolsep{\fill}}llll} \hline\hline
{\rm Decay Modes} &$f_0(\%)$&$f_{\|}(\%)$&$f_{\perp}(\%)$\\ \hline
$B^0_s \to \bar{D}^{*0}(\phi(1020)\to)K^+K^-$ &$54.7^{+1.4+5.2+12.7}_{-2.2-5.4-17.0}$&$34.4^{+1.6+3.7+13.6}_{-1.4-4.0-10.0}$&$10.9^{+0.6+3.1+3.5}_{-0.0-2.2-2.8}$\\
Data~\cite{pdg2018} &$73\pm 15\pm 4$ &$\cdots$ &$\cdots$\\
$B^0_s \to \bar{D}^{*0}(\phi(1680)\to)K^+K^-$ &$45.9^{+1.7+5.9+14.1}_{-1.8-5.3-15.2}$&$39.1^{+1.6+3.9+13.0}_{-1.3-4.3-10.9}$&$14.9^{+0.3+2.7+2.8}_{-0.2-2.6-3.1}$\\
$B^0_{s} \to \bar{D}^{*0} (f_{2}(1270)\to)K^+K^-$&$12.9^{+1.9+7.6+0.9}_{-1.6-5.5-1.4}$&$34.2^{+0.6+2.1+1.5}_{-0.8-3.1-2.8}$&$52.9^{+1.2+3.3+2.0}_{-1.9-4.6-1.0}$\\
$B^0_{s} \to \bar{D}^{*0} (f^{\prime}_{2}(1525)\to)K^+K^-$&$38.8^{+1.9+14.1+11.8}_{-1.7-12.9-12.2}$&$29.1^{+0.8+6.1+6.6}_{-0.9-6.7-5.9}$&$32.1^{+0.9+6.7+5.7}_{-1.2-7.5-5.9}$\\
$B^0_{s} \to \bar{D}^{*0} (f_{2}(1750)\to)K^+K^-$&$34.7^{+1.5+13.6+10.7}_{-1.6-12.2-11.1}$&$31.6^{+0.8+5.8+5.8}_{-0.8-6.7-5.4}$&$33.8^{+0.7+6.2+5.2}_{-0.9-7.2-5.6}$\\
$B^0_{s} \to \bar{D}^{*0} (f_{2}(1950)\to)K^+K^-$&$23.5^{+2.7+11.4+14.0}_{-3.8-9.2-10.3}$&$35.8^{+1.5+4.3+5.5}_{-1.2-5.4-7.1}$&$40.7^{+2.3+5.0+4.8}_{-1.5-6.0-7.0}$\\
$B^0 \to \bar{D}^{*0} (f_{2}(1270)\to)K^+K^-$&$13.4^{+1.8+7.6+9.2}_{-1.7-5.7-7.5}$&$37.8^{+1.0+2.5+3.2}_{-1.0-3.3-3.9}$&$48.8^{+0.7+3.3+4.3}_{-0.8-4.3-5.3}$\\
$B^0 \to \bar{D}^{*0} (f^{\prime}_{2}(1525)\to)K^+K^-$&$16.1^{+2.2+8.9+8.4}_{-2.5-6.7-7.1}$&$32.2^{+1.1+2.6+3.1}_{-0.9-3.4-3.2}$&$51.7^{+1.4+4.2+4.1}_{-1.4-5.5-5.1}$\\
\hline
$B^0_s \to D^{*0}(\phi(1020)\to)K^+K^-$ &$79.6^{+1.6+5.2+1.9}_{-2.8-12.6-6.2}$&$13.5^{+2.3+8.1+2.1}_{-3.1-3.6-3.2}$&$6.9^{+1.5+4.9+4.2}_{-0.0-2.2-0.0}$\\
$B^0_s \to D^{*0}(\phi(1680)\to)K^+K^-$ &$75.6^{+1.8+9.5+1.9}_{-1.2-14.5-11.2}$&$8.9^{+0.6+5.3+8.2}_{-0.4-3.5-0.4}$&$15.5^{+0.8+9.2+3.0}_{-1.4-6.0-1.8}$\\
$B^0_{s} \to D^{*0} (f_{2}(1270)\to)K^+K^-$&$84.2^{+0.2+6.7+1.8}_{-0.3-8.8-2.8}$&$9.1^{+0.2+5.1+1.4}_{-0.0-3.8-1.1}$&$6.7^{+0.1+3.7+1.5}_{-0.2-2.8-0.9}$\\
$B^0_{s} \to D^{*0} (f^{\prime}_{2}(1525)\to)K^+K^-$&$92.6^{+0.7+3.4+2.5}_{-0.3-4.7-1.7}$&$3.9^{+0.2+2.6+0.8}_{-0.3-1.7-1.1}$&$3.5^{+0.3+2.1+0.7}_{-0.5-1.7-1.4}$\\
$B^0_{s} \to D^{*0} (f_{2}(1750)\to)K^+K^-$&$95.5^{+0.8+2.1+1.5}_{-0.3-2.9-2.2}$&$2.2^{+0.3+1.3+2.3}_{-1.1-1.1-3.6}$&$2.3^{+0.3+1.6+0.9}_{-0.2-1.0-0.3}$\\
$B^0_{s} \to D^{*0} (f_{2}(1950)\to)K^+K^-$&$83.2^{+0.6+7.1+3.6}_{-0.5-9.0-1.8}$&$9.1^{+0.6+5.1+0.9}_{-0.3-3.8-1.6}$&$7.6^{+0.1+4.1+0.8}_{-0.2-3.1-1.8}$   \\
$B^+ \to D^{*+} (f_{2}(1270)\to)K^+K^-$&   $79.3^{+0.1+8.3+1.1}_{-0.1-10.7-0.6}$&$11.9^{+0.1+6.2+1.1}_{-0.1-4.8-1.0}$&$8.8^{+0.1+4.6+0.1}_{-0.0-3.5-0.7}$\\
$B^+ \to D^{*+} (f^{\prime}_{2}(1525)\to)K^+K^-$&$72.9^{+2.3+10.3+1.0}_{-1.4-12.4-1.7}$ &$16.1^{+1.0+7.5+1.5}_{-1.2-6.1-0.0}$&$11.0^{+0.4+5.0+1.4}_{-1.2-4.2-2.4}$\\
$B^0 \to D^{*0} (f_{2}(1270)\to)K^+K^-$&$75.4^{+1.0+9.5+4.0}_{-0.7-11.8-1.2}$&$14.2^{+0.4+6.8+0.8}_{-0.2-5.5-2.0}$&$10.4^{+0.3+5.0+0.3}_{-0.8-4.0-2.0}$\\
$B^0 \to D^{*0} (f^{\prime}_{2}(1525)\to)K^+K^-$&$90.8^{+0.7+4.2+3.6}_{-0.9-5.6-2.7}$&$3.0^{+0.5+1.8+1.5}_{-0.0-1.3-1.7}$&$6.2^{+0.6+3.8+1.2}_{-0.5-2.8-2.3}$   \\
\hline\hline
\end{tabular*}
\end{table}

\begin{table}  
\caption{ Branching fractions with (and without) the $r_D^2$ dependent terms in the $B^0_s \to \bar{D}^{*0} (\phi(1020) \to) K^+K^-$ decays.
"FE" and "NFE" represent the contributions from factorizable emission and non-factorizable emission diagrams, respectively. }
\label{Tab:r2}
\begin{center}
\begin{tabular*}{16cm}{@{\extracolsep{\fill}}l l l l ll l l}
\hline\hline
\multicolumn{1}{c}{}&\multicolumn{3}{c}{ FE $\mathcal {B}$ (in $10^{-7}$)}
&\multicolumn{3}{c}{ NFE $\mathcal {B}$ (in $10^{-5}$)} &\multicolumn{1}{c}{\rm Total (in $10^{-5}$)} \\
 {Modes} & $\rm BR_L$  &$\rm BR_N$ &$\rm BR_T$ &$\rm BR_L$  &$\rm BR_N$ &$\rm BR_T$ &\rm Br \\ \hline\hline
$B^0_s \to \bar{D}^{*0} (\phi(1020) \to) K^+K^-$  &0.26 &1.32  &0.04  &1.35  &0.66  &0.24  &2.26\\
$B^0_s \to \bar{D}^{*0} (\phi(1020) \to) K^+K^-(r_D^2)$  &0.27 &1.04  &0.16  &0.95  &0.53  &0.18  &1.76\\
\hline\hline
\end{tabular*}
\end{center}
\end{table}

By using the Eqs.~(\ref{eq:br}-\ref{pol}), the decay amplitudes in Sec.~\ref{sec:3} and Appendix and all the input quantities,
the resultant branching ratios $\mathcal{B}$ and the polarization fractions $f_{\lambda}$ together with the available experimental measurements for the
considered quasi-two-body decays $B_{(s)}\to [ D^{(*)},\bar D^{(*)}](R\to)KK$ are summarized in Tables~\ref{Tab:Db}-\ref{Tab:pola}.
In our numerical calculations for the branching ratios and polarization fractions,
the first theoretical uncertainty results from the parameters of the wave functions of the initial and
final states, such as the shape parameter $\omega_B=0.40\pm0.04$~GeV or $\omega_{B_s}=0.50\pm0.05$~GeV and $C_{D}=0.5\pm 0.1$, and the decay constant $f_{B_{(s)}}$ in $B_{(s)}$ meson wave function.
The second one is due to the Gegenbauer moments in DAs of $KK$ pair with different intermediate resonances, which are supposed to be varied with a $20\%$ range for the error estimation.
With the improvements of the experiments and the deeper theoretical developments, this kind of uncertainties will be reduced.
The last one is caused by the variation of the hard scale $t$ from $0.75t$ to $1.25t$ (without changing $1/b_i$) and the QCD scale $\Lambda_{\rm QCD}=0.25\pm0.05$~GeV, which characterizes the effect of the next-to-leading-order QCD contributions.
The possible errors due to the uncertainties of $m_c$ and CKM matrix elements are very small and can be neglected safely.

Compared with $B_{(s)}\to D^{(*)}R\to D^{(*)}KK$ decays,
the $B_{(s)} \to \bar{D}^{(*)}R\to \bar{D}^{(*)}KK$ ones are enhanced by the CKM matrix
elements $|V_{cb}/V_{ub}|^2$, especially for those without a strange quark in the four-quark operators.
So for most of the $B_{(s)} \to \bar{D}^{(*)}R\to \bar{D}^{(*)}KK$
decays, the branching ratios are at the order
from $10^{-7}$ to $10^{-5}$; while for the $B_{(s)}\to D^{(*)}R\to D^{(*)}KK$ decays, the branching ratios are at the order from $10^{-9}$ to $10^{-7}$.
The two-body branching fraction ${\cal B}(B \to D^{(*)}R)$ can be extracted from the corresponding quasi-two-body decay modes in
Tables~\ref{Tab:Db}-\ref{Tab:Ds} under the narrow width approximation relation
\begin{eqnarray}
\mathcal{B}(B \to D^{(*)}R\to  D^{(*)} K^+ K^- ) = \mathcal{B}( B \to D^{(*)}R) \cdot {\mathcal B}(R \to K^+K^-).\label{eq:def1}
\end{eqnarray}
By using the measured branching fractions ${\mathcal B}(\phi(1020) \to K^+K^-)=49.2\%$, ${\mathcal B}(f_2(1270) \to K^+K^-)=2.3\%$
and ${\mathcal B}(f_2^{\prime}(1525) \to K^+K^-)=44.4\%$~\cite{pdg2018} as an input, we can extract the branching ratios of two-body decays
$B \to D^{(*)}\phi(1020)$, $B \to D^{(*)}f_2(1270)$ and $B \to D^{(*)}f_2^{\prime}(1525)$, which agree with those from the two-body analyses
based on the PQCD approach~\cite{prd78-014018,jpg37-015002,prd86-094001} within errors.
The tiny differences between our predictions of the two-body decays and previous results of Refs.~\cite{prd78-014018,jpg37-015002,prd86-094001}
are mainly due to the parametric origins and the power corrections related to the ratio $r_D^2=m^2_D/m^2_B$.
As we know, a significant impact of nonfactorizable contribution is expected for a colour suppressed decay mode.
Taking the decay channel $B^0_s \to \bar{D}^{*0} (\phi(1020) \to) K^+K^-$ as an example, one can see that the inclusion of the power correction $r_D^2$ can
suppress the branching ratio efficiently, especially for the contributions of nonfactorizable emission diagrams as shown in Table~\ref{Tab:r2}.

Now, we come to discuss the contributions of $P$-wave resonances.
It is obvious that the PQCD predictions of the ${\cal B}(B^0_s \to \bar{D}^{(*)0} \phi(1020) \to \bar{D}^{(*)0} K^+K^-)$ are
consistent well with the experimental data within errors.
Taking $B^0_s \to \bar D^{0}K^+K^-$ decay as an example, we can calculate the total $P$-wave resonance contributions
${\cal B}(B^0_s \to \bar D^{0}K^+K^-)_{\text{P-wave}}=1.32 \times 10^{-5}$, which is nearly equal to the sum of resonance contributions
from $\phi(1020)$ and $\phi(1680)$.
The main reason is that the interference between the two $P$-wave resonant states $\phi(1020)$ and $\phi(1680)$ is really small due to
the rather narrow width of the former $(\Gamma_{\phi(1020)}=4.25 {\rm MeV})$.
Since the contribution from high mass resonance is one order of magnitude smaller, the $P$-wave resonance contribution is dominant by $\phi(1020)$.
From the numerical results given in Table~\ref{Tab:Db} and~\ref{Tab:Dsb}, we obtain the relative ratio $R(\phi)$ between the branching ratio of $B^0_s$
meson into $\bar{D}^{*0}$ and $\bar{D}^{0}$ plus the resonance $\phi$,
\begin{eqnarray}
R^{\text{PQCD}}(\phi)&=&\frac{{\cal B}(B^0_s \to \bar{D}^{*0}(\phi\to) K^+K^-)}{{\cal B}(B^0_s \to \bar{D}^{0}(\phi\to) K^+K^-)}
= 1.45^{+1.28}_{-1.06},
\end{eqnarray}
which is in agreement with  the corresponding ratio of $\phi$ reported by the LHCb measurement~\cite{prd98-071103},
\begin{eqnarray}
R^{\text{exp}}(\phi)&=&\frac{{\cal B}(B^0_s \to \bar{D}^{*0}(\phi\to) K^+K^-)}{{\cal B}(B^0_s \to \bar{D}^{0}(\phi\to) K^+K^-)}=1.23\pm0.20\pm0.08,
\end{eqnarray}
where the first uncertainty is statistical and the second is systematic.

Utilizing the PQCD prediction in Table~\ref{Tab:Db} and $\mathcal{B}(B^0_s\to\bar{D}^{0}\bar{K}^{*0}))=(2.33^{+1.18}_{-0.69})
\times 10^{-4}$ taken from our previous work in Ref.~\cite{epjc79-539} together with the two known branching ratios
$\mathcal{B}(\phi \to K^+K^-)=(49.2 \pm 0.5)\%$ and $\mathcal{B}(K^* \to K\pi)\sim 100\%$~\cite{pdg2018} , we expect that
\begin{eqnarray}\label{eq:rr}
R^{\text{PQCD}}_{\phi/\bar{K}^{*0}}=\frac{\mathcal{B}(B^0_s \to \bar{D}^{0}\phi)}{\mathcal{B}(B^0_s \to \bar{D}^{0}\bar{K}^{*0})}=0.11^{+0.09}_{-0.07},
\end{eqnarray}
which is a bit larger than the data measured by LHCb~\cite{plb727-403},
\begin{eqnarray}\label{eq:rr1}
R^{\text{exp}}_{\phi/\bar{K}^{*0}}=\frac{\mathcal{B}(B^0_s \to \bar{D}^{0}\phi)}{\mathcal{B}(B^0_s \to \bar{D}^{0}\bar{K}^{*0})}
=0.069\pm0.013(\text{stat})\pm0.007(\text{syst}).
\end{eqnarray}
Recalling that the theoretical errors are relatively large, one still can count them as being consistent within one standard deviation.

In contrast to the vector resonances, the identification of the scalar mesons is a long-standing puzzle.
Scalar resonances are difficult to resolve because some of them have large decay widths, which cause a strong overlap between resonances and background.
The prominent appearance of the $f_0(980)$ implies a dominant ($\bar{s}s$) component in the semileptonic $D_s$ decays and decays of $B_{(s)}$ mesons.
Ratios of decay rates of $B$ and/or $B_s$ mesons into $J/\psi$
plus $f_0(980)$ or $f_0(500)$ were proposed to allow for an extraction of
the flavor mixing angle and to probe the tetraquark nature of those
mesons within a certain model~\cite{epjc71-1832,prl111-062001}.
The phenomenological fits of the LHCb collaboration do neither allow
for a contribution of the $f_0(980)$ in the $B\to J/\psi\pi\pi$~\cite{prd90-012003} nor an $f_0(500)$ in $B_s \to J/\psi\pi\pi$ decays~\cite{prd89-092006} by using the isobar model.
The authors conclude that their data is incompatible with a model where $f_0(980)$ is formed from two quarks and two antiquarks (tetraquarks) at the eight standard deviation level.
In addition, they extract an upper limit for the mixing angle of $17^\circ$ at $90\%$ confidence level between
the $f_0(980)$ and the $f_0(500)$ that would correspond to a substantial
($\bar{s}s$) content in $f_0(980)$~\cite{prd90-012003}.
However, a substantial $f_0(980)$ contribution is also found in the $B$-decays in a dispersive analysis of the same data that allows for a model-independent inclusion of the hadronic
final state interactions in Ref.~\cite{jhep1602-009}, which puts into question the
conclusions of Ref.~\cite{prd90-012003}.
At this stage, the quark structure of scalar particles are still quite controversial.
As a first approximation, the $S$-wave time-like form factor $F_S(m^2_{KK})$
used to parameterize the $S$-wave two-kaon distribution amplitude has been determined in Ref.~\cite{epjc79-792}.
Therefore, we take into account $f_0(980)$ and $f_0(1370)$ in the $\bar{s}{s}$ density operator.

For the $S$-wave resonances $f_0(980)$ and $f_0(1370)$, firstly we define the ratio between the $f_0(980) \to K^+K^-$ and $f_0(980) \to \pi^+\pi^-$,
\begin{eqnarray}\label{eq:rr2}
R_{K/\pi}=\frac{\mathcal{B}(B^0_s\rightarrow \bar{D}^0 f_0(980)(\rightarrow K^+K^-))}{\mathcal{B}(B^0_s\rightarrow \bar{D}^0 f_0(980)(\rightarrow \pi^+\pi^-))}
\approx \frac{\mathcal{B}( f_0(980)\rightarrow K^+K^-)}{\mathcal{B}(f_0(980)\rightarrow \pi^+\pi^-)}.
\end{eqnarray}
On the experimental side, $BABAR$ measured
the ratio of the partial decay width of $f_0(980)\rightarrow K^+K^-$ to $f_0(980)\rightarrow \pi^+\pi^-$
of $R^{exp}_{K/\pi}=0.69\pm 0.32$ using $B\rightarrow KK^+K^-$ and $B\rightarrow K\pi^+\pi^-$ decays~\cite{prd74-032003}.
Meanwhile, BES performs a partial wave analysis of $\chi_{c0}\rightarrow f_0(980)f_0(980)\rightarrow \pi^+\pi^-\pi^+\pi^-$
and $\chi_{c0}\rightarrow f_0(980)f_0(980)\rightarrow \pi^+\pi^-K^+K^-$ in $\psi(2S)\rightarrow \gamma \chi_{c0}$ decay
and  extracts the ratio as $R^{exp}_{K/\pi}=0.25^{+0.17}_{-0.11}$~\cite{prd70-092002,prd72-092002}.
Their average yields $R^{exp}_{K/\pi}=0.35^{+0.15}_{-0.14}$~\cite{prd92-032002}.
Recently, LHCb Collaboration~\cite{jhep08-005} reported a measurement,
$\mathcal{B}( B^0_s\rightarrow \bar{D}^0 f_0(980) \to \bar{D}^0 \pi^+\pi^- )=(1.7\pm1.1)\times10^{-6}$, one can roughly infer the wide range of
$\mathcal{B}^{\text{exp}}( B^0_s\rightarrow \bar{D}^0 f_0(980) \to \bar{D}^0 K^+K^- )=(0.2-1.0)\times 10^{-6}$ combining with the ratio $R^{\text{exp}}_{K/\pi}$.
Our PQCD prediction $\mathcal{B}( B^0_s\rightarrow \bar{D}^0 f_0(980) \to \bar{D}^0 K^+K^- )=(1.36^{+1.20}_{-0.96})\times10^{-6}$ lies in above range within errors,
which is  also agree with the upper limit $1.55 \times10^{-6}$ given by PDG~\cite{pdg2018} as shown in Table~\ref{Tab:Db}.
However, BES also measured $R_{K/\pi}=0.625\pm0.21$ by studying the decays $J/\psi\to\phi f_0(980)\to \phi\pi^+\pi^-$
and $J/\psi \to \phi f_0(980)\to \phi K^+K^-$ in Ref.~\cite{plb607-243}.
It seems that we hardly can reach a reliable and universal $R_{K/\pi}$.
It should be stressed that there are large uncertainties in both experimental measurements and the theoretical calculations,
so the discrepancy between the data and the theoretical results could be clarified with the high precision experimental
data and the high precision of theoretical calculations.
Since the situation of the knowledge about the $f_0(1370)$ decaying into $KK$ or $\pi\pi$  is rather unclear, no evidence of the $f_0(1370)$ resonance in $B \to DKK$ decay
has been reported so far.
Our predictions on the $KK$ channel involving the scalar resonances $f_0(980)$ and $f_0(1370)$ in Tables~\ref{Tab:Db}-\ref{Tab:Ds}
will be investigated at the ongoing LHCb and Belle-II experiments in the future.

The branching ratios of the considered $D$-wave resonances are also presented in Tables~\ref{Tab:Db}-\ref{Tab:Ds}.
Belle~\cite{prd76-012006} provided a Dalitz plot analysis of $\bar{B}^0 \to D^0 \pi^+\pi^-$ decays and obtain the branching ratio of
$\mathcal{B}(\bar{B}^0 \to D^0f_2)=(1.20\pm0.18(\text{stat})\pm0.21(\text{syst}))\times 10^{-4}$.
Afterwards, LHCb~\cite{prd92-032002} analysed the resonant substructures of $B^0 \to \bar{D}^0 \pi^+\pi^-$ decays and reported
the branching ratio of $\mathcal{B}(B^0 \to \bar{D}^0f_2)=(1.68\pm0.11(\text{stat})\pm0.21(\text{syst}))\times 10^{-4}\text{(Isobar)}$.
Their weighted average yields $\mathcal{B}(B^0 \to \bar{D}^0f_2)=(1.56\pm0.21)\times 10^{-4}$~\cite{pdg2018}.
For a more direct comparison, we can extract out the branching fraction for the two-body decay
$\mathcal{B}(B^0 \to \bar{D}^0f_2)=(1.54^{+1.08}_{-0.97})\times 10^{-4}$ through the narrow-width approximation
relation in Eq.~(\ref{eq:def1}), which agrees with above experimental results within uncertainties.
Furthermore, we have explicitly written the relative ratio of $\mathcal{B}(B^0_s \to \bar{D}^0 f_2(1270)(\to K^+K^-))$
compared to $\mathcal{B}(B^0_s\to \bar{D}^0 f_2(1270)(\to \pi^+\pi^-))$ in the narrow width limit,
\begin{eqnarray}
R(f_2)=\frac{\mathcal{B}(B^0_s \to \bar{D}^0 f_2(1270)(\to K^+K^-))}{\mathcal{B}(B^0_s\to \bar{D}^0 f_2(1270)(\to \pi^+\pi^-))}\approx
\frac{\mathcal{B}(f_2(1270)\to K^+K^-)}{\mathcal{B}(f_2(1270)\to \pi^+\pi^-)}.
\end{eqnarray}
Thereby, using the experimental data $\mathcal{B}(f_2(1270)\to K^+K^-)=\frac{1}{2}(4.6)\%$,
$\mathcal{B}(f_2(1270)\to \pi^+\pi^-)=\frac{2}{3}(84.2)\%$ from PDG~\cite{pdg2018} and the
measurements $\mathcal{B}(B^0_s\to \bar{D}^0 f_2(1270)\to \bar{D}^0\pi^+\pi^-)=(6.8\pm2.4)\times 10^{-5}$
from Belle~\cite{prd76-012006} and $\mathcal{B}(B^0_s\to \bar{D}^0 f_2(1270)\to \bar{D}^0\pi^+\pi^-)=(9.5\pm1.3)\times 10^{-5}(\text{Isobar})$
from LHCb~\cite{prd92-032002}, the experimental value of $B^0_s\to \bar{D}^0 f_2(1270)(\to K^+K^-)$ decay is estimated to be
\begin{eqnarray}
\mathcal{B}(B^0_s\to \bar{D}^0 (f_2(1270)\to) K^+K^-)=\left\{
\begin{aligned}
&2.7\times 10^{-6}, \quad\quad  &\text{Belle}, \nonumber\\
&3.8\times 10^{-6}, \quad\quad  &\text{LHCb},  \nonumber\\
\end{aligned}\right.
\end{eqnarray}
which is one order of magnitude larger than our prediction $\mathcal{B}(B^0_s\to \bar{D}^0 f_2(1270)\to \bar{D}^0K^+K^-)=1.87\times 10^{-7}$ in its central value.
Since the property of the tensor resonance is not well understood and the theoretical uncertainties are relatively large, this issue needs to be further clarified in the future.

From the numerical results given in Table~\ref{Tab:Db}, we can evaluate the relative branching ratios between two tensor modes,
\begin{eqnarray}
R_{f_2/f^{\prime}_2}=\frac{\mathcal{B}(B_s^0\to \bar{D}^0 f_2(1270))}{\mathcal{B}(B_s^0 \to \bar{D}^0 f^{\prime}_2(1525))}=0.05^{+0.10}_{-0.03},
\end{eqnarray}
which can be tested by the forthcoming LHCb and Belle-II experiments.

\begin{figure}[tbp]
\centerline{
\hspace{1.5cm}\subfigure{\epsfxsize=11 cm \epsffile{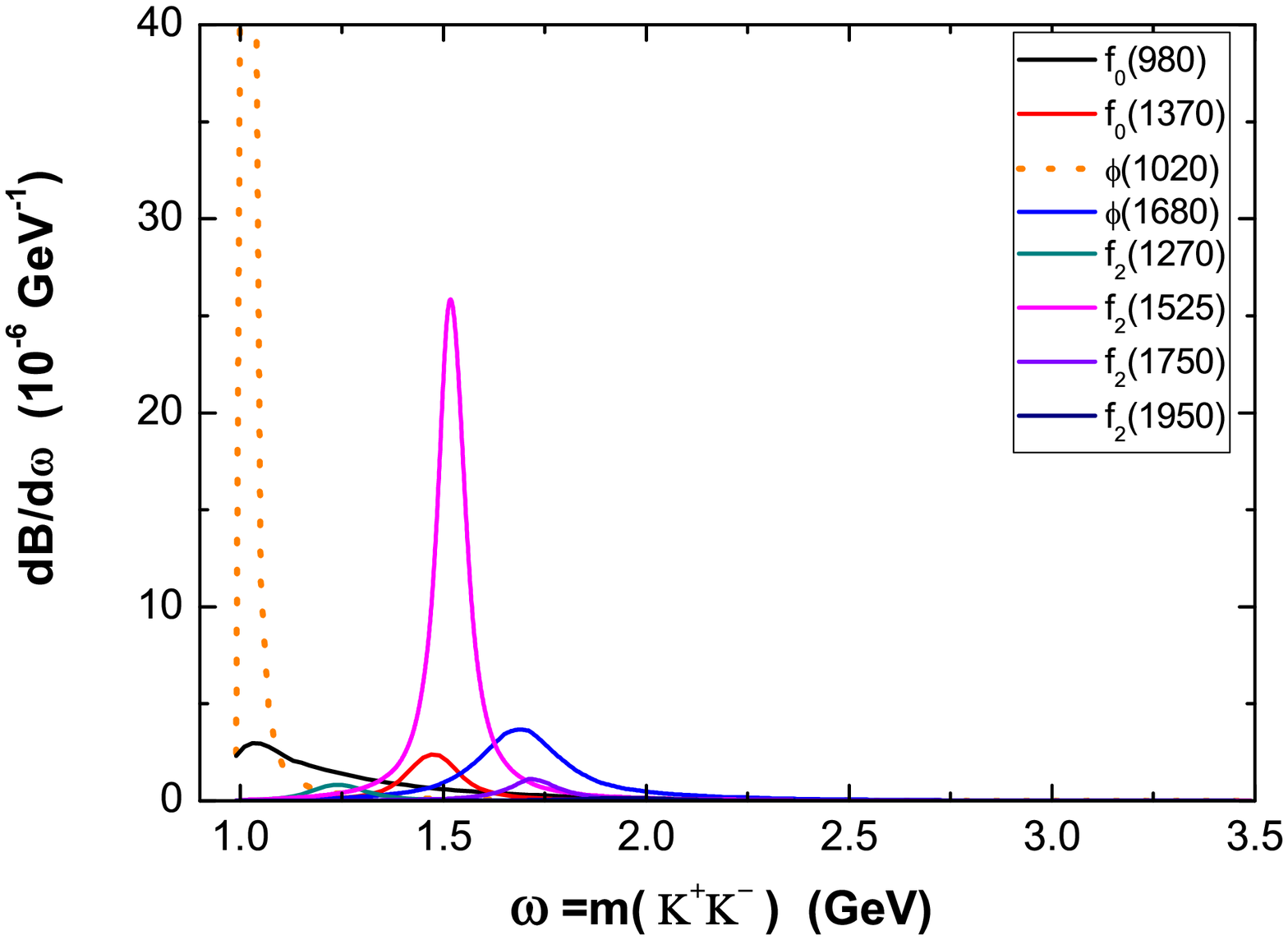} }
\hspace{-3.0cm}\subfigure{ \epsfxsize=11 cm \epsffile{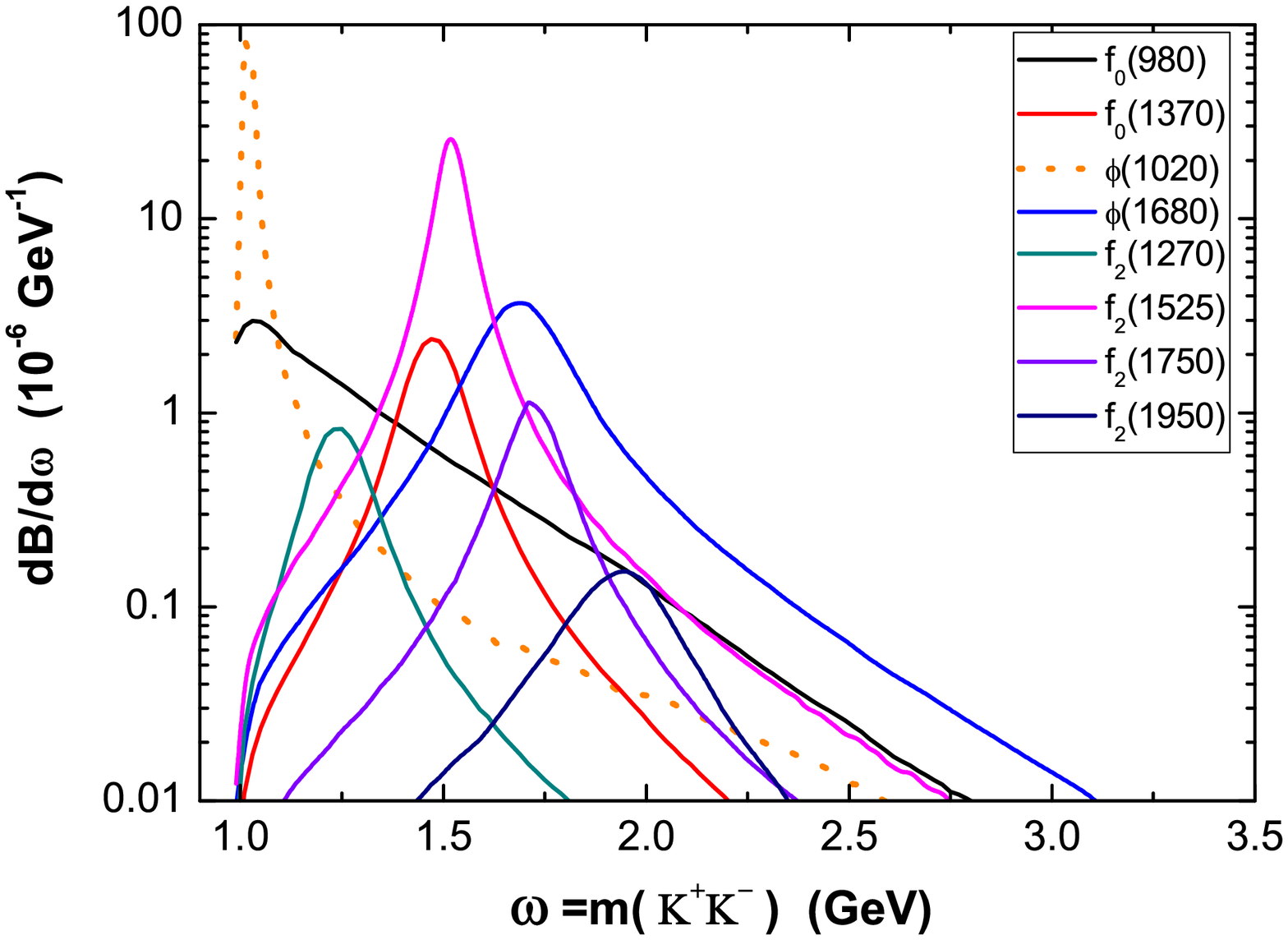}}}
\vspace{-1.5cm}
  {\scriptsize\bf (a)\hspace{8.5cm}(b)}
\caption{(a) The $m(K^+K^-)$-dependence of the differential decay rates $d{\cal B}/d\omega$
for the contributions from the resonances $f_0(980),f_0(1370),\phi(1020),\phi(1680),f_2(1270),f^{\prime}_2(1525),f_2(1750)$ and $f_2(1950)$
to $B^0_s\to \bar{D}^0 K^+K^-$ decay with a linear scale. (b) The same curves are shown with a logarithmic scale.}
\label{fig-br}
\end{figure}

In Fig.~\ref{fig-br}, we show the $\omega$-dependence of the differential decay rate $d{\cal B}(B^0_s \to  \bar{D}^0 K^+K^-)/d\omega$
after the inclusion of the possible contributions from the resonant states $f_0(980)$ (the black solid curve), $f_0(1370)$ (the red solid curve), $\phi(1020)$ (the orange dotted curve), $\phi(1680)$ (the blue solid curve), $f_2(1270)$ (the cyan solid curve), $f_2^{\prime}(1525)$ (the magenta solid curve),  $f_2(1750)$ (the violet solid curve) and $f_2(1950)$ (the navy solid curve).
To see clearly all the peaks of the various resonances, we draw them both in Fig.~\ref{fig-br}(a) with a linear scale and Fig.~\ref{fig-br}(b) with a logarithmic one.
For the considered decay mode $B^0_s \to  \bar{D}^0 K^+K^-$, the dynamical limit on the value of invariant mass $\omega$ is $(m_{K^+}+m_{K^-}) \leq \omega \leq (m_B-m_D)$.
It is clear that an appreciable peak arising from the $\phi(1020)$ resonance, followed by $f_2^{\prime}(1525)$.
Another three resonance peaks of $f_0(980)$, $f_0(1370)$, and $\phi(1680)$ have relatively smaller strength than the $f_2^{\prime}(1525)$ one,
while their broader widths compensate the integrated strength over the whole phase space.
Therefore, the branching ratios of the four components are of a comparable size as predicted in our work.
The contribution of the tensor $f_2(1270)$ is really small in $B^0_s \to  \bar{D}^0 K^+K^-$ decay mode, since it decays predominantly into $\pi\pi$.
Apart from above obvious signal peak, there are two visible structures at about 1750 MeV and 1950 MeV
in Fig.~\ref{fig-br}(b).
Obviously, the differential branching ratios of these decays exhibit peaks at the pole mass of the resonant states.
Thus, the main portion of the branching ratio lies in the region around the resonance as expected.
For $B^0_s \to \bar{D}^0 \phi \to \bar{D}^0 K^+K^-$ decay, the central values of the branching ratio ${\cal B}$ are $5.82\times 10^{-6} $ and $8.29\times 10^{-6}$
when the integration over $\omega$ is limited in the range of
$\omega=[m_{\phi}-0.5\Gamma_{\phi}, m_{\phi}+0.5\Gamma_{\phi}]$ or
$\omega=[m_{\phi}-\Gamma_{\phi}, m_{\phi}+\Gamma_{\phi}]$ respectively, which amount to
$48.1\%$ and $68.5\%$ of the total branching ratio ${\cal B}=1.21\times10^{-5}$ as listed in Table~\ref{Tab:Db}.

The $B \to (\bar{D}^{*},D^{*})[\phi,f_2^{(\prime)} \to]K^+K^-$ decays are vector-vector (vector-tensor) modes and can proceed through different polarization amplitudes.
We display the polarization fractions associated with the available data in Table~\ref{Tab:pola},
which have the same origin of theoretical uncertainties as the branching ratios.
It is easy to see that the fraction of the longitudinal polarization can be generally reduced to about $\sim 50\%$,
while the parallel and perpendicular ones are roughly equal.
The results are quite different from the expectation in the factorization assumption that the longitudinal
polarization should dominate based on the quark helicity analysis~\cite{zpc1-269,prd64-117503}.
For $B^0_s \to \bar{D}^{*0} (\phi(1020)\to)K^+K^-$ decay, although the central value of
$f_0\approx 55\%$ is a little smaller than the measured one  $f_0^{exp}=73\%$, but they are consistent with each other
due to the still large theoretical and experimental uncertainties.

It is worthwhile to stress that the polarization fractions of the colour-suppressed decays $B\to \bar D^*(f_2\to)KK$ and $B\to D^*(f_2\to)KK$
are quite different from each other.
For $B\to \bar D^*(f_2\to)KK$ ($\bar b\to \bar c$ transition) decays, the percentage of the transverse polarizations ($f_\|+f_\bot$) can be as large
as $80\%$, while for $B\to D^*(f_2\to)KK$ ($\bar b\to \bar u$ transition) decays, the percentage of the transverse polarizations are only
at the range of $10\%\sim 20\%$.
This situation can be understood as the following reasons~\cite{prd86-094001}:
The $\bar D^*$ meson in $B\to \bar D^*(f_2\to)KK$ decays is longitudinal polarized since the $\bar c$ quark and the $u$
quark  in the $\bar D^*$ meson produced through the $(V-A)$ current are right-handed and left-handed, respectively.
Because the $\bar c$ quack is massive, the helicity of the $\bar c$ quark can flip easily from right handed to left handed.
Thus, the $\bar D^*$ meson can be transversely polarized with the polarization $\lambda=-1$.
The recoiled tensor meson can also be transversely polarized with the polarization $\lambda=-1$ due to the contribution of orbital angular momentum.
Then the transverse polarized contribution in $B\to \bar D^*(f_2\to)KK$ decays can be sizable.
While for the $D^*$ meson, the $\bar u$ quark in the $D^*$ meson is right handed and the $c$ quark can flip from
left handed to right handed, which makes the $D^*$ meson transversely polarized with the polarization $\lambda=+1$.
In order to have the same polarization with the $D^*$ meson, the recoiled transversely polarized tensor meson needs contributions from both
orbital angular momentum and spin, so the situation is symmetric.
But the wave function of the tensor meson is asymmetric.
Therefore, the transversely polarized contribution in $B\to D^*(f_2\to)KK$ is suppressed on account of Bose statistics.
More precise measurements of such decay channels are expected to help us to test and improve our theoretical calculations.

\section{CONCLUSION}\label{sec:5}

In this work, we have performed an systematic analysis of various partial wave resonant contributions to three-body $B_{(s)} \to [ D^{(*)}, \bar{D}^{(*)}] KK$ decays in the PQCD approach.
For $B^0_s \to [ D^{(*)}, \bar{D}^{(*)}] K^+K^-$ decay modes,
the dominant contributions are expected to be from $P$-wave resonance $\phi(1020)$, $D$-wave resonance $f_2^{\prime}(1525)$.
Besides the two prominent components mentioned above, some significant resonant structures also exhibit in the
$K^+K^-$ invariant mass spectrum, like $f_0(980)$, $f_0(1370)$ and $\phi(1680)$.
Such resonances have relatively smaller strength than the $f_2^{\prime}(1525)$ resonance, but their broader widths compensate the integrated strength over the entire phase space.
Since $\pi\pi$ is the dominant decay mode of $f_2(1270)$, the contribution of the tensor $f_2(1270)$ is indeed small.
The whole pattern of $B_{(s)} \to [ D^{(*)}, \bar{D}^{(*)}] KK$ decays will be confronted with the experimental data in the future.

Each partial wave contribution can be parameterized into the corresponding time-like form factor, which contains the final state interactions
between the kaons in the resonant regions.
The Flatt\'e model for the $f_0(980)$ resonance and the Breit-Wigner formulas for other resonances have been adopted to
parameterize the time-like form factors $F_{S,P,D}(\omega^2)$ involved in the dikaon DAs.
Using the determined parameters of the two-kaon DAs in our previous work,
we have predicted the branching ratios of $B_{(s)} \to [ D^{(*)}, \bar{D}^{(*)} ] (R\to)KK$ decay channels.
It has been shown that our predictions of the branching ratios for most of the considered decays are in good agreement with the existing data within the errors.
The branching ratios of the two-body $B_{(s)} \to [ D^{(*)}, \bar{D}^{(*)}] R$ can be extracted from the corresponding quasi-two-body modes
by employing the narrow width approximation.
Moreover, we calculated the polarization fractions of the vector-vector and vector-tensor decay modes in detail.
For most of the considered channels, the transverse polarization has been found to be of the similar size as the longitudinal one,
which is quite different from the general expectation in the factorization assumption.
More precise data from the LHCb and the future Belle II will test our predictions.

\begin{acknowledgments}
Many thanks to Hsiang-nan Li for valuable discussions.
This work was supported by the National Natural Science Foundation of China under the No.~11947013, No.~11605060, No.~11775117, and No.~11547020.
Y.~Li  is also supported by the Natural Science Foundation of Jiangsu Province under Grant No.~BK20190508 and the Research Start-up Funding of Nanjing
Agricultural University.
Z.~Rui is supported in part by the Natural Science Foundation of Hebei Province under Grant No.~A2019209449.

\end{acknowledgments}

\appendix\label{sec:ap}
\section{Decay amplitudes}
For $B_{(s)}\to \bar D^{(*)}(R\to)KK$ decays (R denotes the various partial wave resonances ), the expressions of the individual amplitudes $F^{LL}_{e(a)}$ and $ M^{LL}_{e(a)}$  can be straightforwardly obtained by evaluating the Feyman diagrams in Fig.~\ref{fig:fig1}.
Performing the standard PQCD calculations, one gets the following expressions of the relevant amplitudes:
\begin{enumerate}
\item[$\bullet$]  $B_{(s)}\to \bar D(f_0\to)KK$
\begin{eqnarray}
F_{ef_0}^{LL}&=&\frac{8\pi C_Fm^4_Bf_D}{\sqrt{\eta-\eta r_D^2}}\int^1_0 dx_B dz \int_0^{1/\Lambda} bdb b_Bdb_B\phi_B(x_B,b_B)\non
     &&\times \Big \{ \Big[\phi_0(z) \sqrt{\eta -\eta  r_D^2} (r_D^2 (-2 \eta  (z+1)+2 z+1)+(\eta -1) (z+1))\non
     &&+\eta  (1-r_D^2) \left ( (\phi_t(z)+\phi_s(z))(1+2z(r_D^2 -1))(\eta -1)+r_D^2(\phi_t(z)-\phi_s(z) ) \right ) \Big]\non
     && \cdot E_e(t_a)h_a(x_B,z,b,b_B)S_t(z)\non
     && +\sqrt{\eta -\eta  r_D^2} \Big [\phi_0(z)\left  ( r_D^2 (\eta ^2-x_B)-(\eta -1) \eta \right )+2 \phi_s(z) \sqrt{\eta -\eta  r_D^2} \non
     && \cdot \left (\eta +r_D^2 (-2 \eta+x_B+1)-1\right ) \Big ] \cdot E_e(t_b)h_a(x_B,z,b_B,b)S_t(|x_B-\eta|)\Big \},
\end{eqnarray}
\begin{eqnarray}
F_{af_0}^{LL}&=&\frac{8\pi C_Fm^4_Bf_B}{\sqrt{\eta-\eta r_D^2}}\int^1_0 dx_3 dz \int_0^{1/\Lambda} bdb b_3db_3\phi_D(x_3,b_3)\non
     &&\times \Big \{[\phi_0(z) \sqrt{\eta -\eta  r_D^2} (r_D^2 (2 \eta -2 (\eta -1) z-1)+(\eta -1) (z-1))-2 \eta \non
     && \times r_D (r_D^2-1)^2 z (\phi_t(z)-\phi_s(z))+4 \eta  r_D(r_D^2-1)\phi_s(z) ]E_a(t_e)\non
     &&\times h_e(z,x_3,b,b_3)S_t(z)+ \sqrt{\eta -\eta  r_D^2}[\phi_0(z) (\eta  (\eta -\eta  r_D^2-1)+(\eta -1)^2 \non
     &&\times(r_D^2-1) x_3)+2 r_D \phi_s(z) \sqrt{\eta -\eta  r_D^2} (\eta -r_D^2-\eta  x_3+x_3+1)]\non
     &&\times E_a(t_f)h_f(z,x_3,b_3,b)S_t(|\eta(x_3-1)-x_3|)\Big \},
\\
\non
M_{ef_0}^{LL}&=&\frac{32\pi C_Fm^4_B}{\sqrt{6(\eta-\eta r_D^2)}}\int^1_0 dx_B dzdx_3\int_0^{1/\Lambda} b_3db_3 b_Bdb_B\phi_B(x_B,b_B)\phi_D(x_3,b_3)\non
     &&\times \Big \{[\phi_0(z) (-\eta +r_D^2+1) \sqrt{\eta -\eta  r_D^2} ((1-r_D^2)((x_3+x_B-1)-(x_3+z)\eta) \non
     &&+\eta  (1-2r_D^2))+\eta  (-r_D^2+1)( \phi_t(z) ((\eta -1) z(1-r_D^2)-r_D^2 ((1-\eta)  x_3  \non
     &&+x_B))+ \phi_s(z) (r_D^2 ( x_3+2+z)-z)(\eta -1)] E_n(t_c)h_c(x_B,z,x_3,b_B,b_3)\non
     &&-[\phi_0(z) \sqrt{\eta -\eta  r_D^2} (-\eta +(2 \eta -1) r_D^2+1) ((r_D^2-1) z+(\eta -1) x_3+x_B)\non
     &&+\eta  (-r_D^2+1)(( \phi_t(z) -\phi_s(z))r_D^2 ((\eta -1) x_3+x_B)+(\eta -1) z(r_D^2-1)\non
     &&\times ( \phi_t(z)+\phi_s(z)))]E_n(t_d)h_d(x_B,z,x_3,b_B,b_3)\Big \},
\end{eqnarray}
\begin{eqnarray}
M_{af_0}^{LL}&=&\frac{32\pi C_Fm^4_B}{\sqrt{6(\eta-\eta r_D^2)}}\int^1_0 dx_B dzdx_3\int_0^{1/\Lambda} bdb b_Bdb_B\phi_B(x_B,b_B)\phi_D(x_3,b_3)\non
     &&\times \Big \{ [\phi_0(z) \sqrt{\eta -\eta  r_D^2} (r_D^2 (\eta  (-\eta  (x_3+z-2)+x_3+x_B)-1)+(\eta -1) (\eta  (x_3\non
     &&+z-1)-x_3-x_B))+\eta  r_D (1-r_D^2)((\phi_t(z)-\phi_s(z) ) ((1-\eta)(x_3-1) +x_B) \non
     &&+(\phi_t(z)+\phi_s(z) )z-4\phi_s(z))]E_n(t_g)h_g(x_B,z,x_3,b,b_B)\non
     &&-[\phi_0(z) (-\eta +r_D^2-1) \sqrt{\eta -\eta  r_D^2} (r_D^2 (\eta  (x_3+z-2)-x_3+x_B-z+1)\non
     &&+( \eta-1) (1- z))+\eta  r_D(1-r_D^2) ((\phi_t(z)-\phi_s(z) )  (r_D^2-1)(z-1) +(\phi_t(z) \non
     &&+\phi_s(z) )(\eta-1) x_3+x_B-\eta)] E_n(t_h)h_h(x_B,z,x_3,b,b_B)\Big \}.
\end{eqnarray}
\item[$\bullet$]  $B_{(s)}\to \bar D^*(f_0\to)KK$
\begin{eqnarray}
F_{ef_0}^{LL}&=&-\frac{8\pi C_Fm^4_Bf_{D^*}}{\sqrt{(\eta -\eta  r_{D}^2)(1-\eta)}}\int^1_0 dx_B dz \int_0^{1/\Lambda} bdb b_Bdb_B\phi_B(x_B,b_B)\non
     &&\times \Big \{[\phi_0(z) \sqrt{\eta -\eta  r_{D}^2} (r_{D}^2 (1-2 (\eta -1) z)+(\eta -1) (z+1))\non
     &&+\eta  (-r_{D}^2+1)( (\phi_s(z)+\phi_t(z))(1+2z(r_{D}^2 -1))(\eta -1)+r_{D}^2(-\phi_t(z)+\phi_s(z))) ] \non
     && \cdot E_e(t_a)h_a(x_B,z,b,b_B)S_t(z) - \sqrt{\eta -\eta  r_{D}^2} \big [\phi_0(z) (-(r_{D}^2 ((\eta -2) \eta +x_B)-(\eta -1) \eta ))\non
     &&+2 \phi_s(z) \sqrt{\eta -\eta  r_{D}^2} (-\eta +r_{D}^2 (x_B-1)+1) \big] \cdot E_e(t_b)h_a(x_B,z,b_B,b)S_t(|x_B-\eta|)\Big \}, \non
\\
\non
M_{ef_0}^{LL}&=&\frac{32\pi C_Fm^4_B}{\sqrt{6(\eta-\eta r_{D}^2)(1-\eta)}}\int^1_0 dx_B dzdx_3\int_0^{1/\Lambda} b_3db_3 b_Bdb_B\phi_B(x_B,b_B)\phi_D(x_3,b_3)\non
     &&\times \Big \{[\phi_0(z) (\eta +r_{D}^2-1) \sqrt{\eta -\eta  r_{D}^2} ((\eta(1-x_3-z) +r_{D}^2 \eta  (x_3+z-2)\non
     &&+(1-r_{D}^2)(x_3+x_B-1))+\eta  (-r_{D}^2+1)( \phi_s(z) ((\eta -1) z(1-r_{D}^2)-r_{D}^2 ((1  \non
     &&-\eta)  x_3+x_B))+ \phi_t(z) (r_{D}^2 ( x_3+2+z)-z)(\eta -1)] \cdot  E_n(t_c)h_c(x_B,z,x_3,b_B,b_3)\non
     &&-[\phi_0(z) (\eta +r_{D}^2-1) \sqrt{\eta -\eta  r_{D}^2} ((r_{D}^2-1) z+(\eta -1) x_3+x_B)\non
     &&+\eta  (-r_{D}^2+1)(( \phi_t(z)-\phi_s(z)) r_{D}^2 ((\eta -1) x_3+x_B)+( \phi_t(z)\non
     && +\phi_s(z))z(1-\eta)(r_{D}^2-1) ] \cdot E_n(t_d)h_d(x_B,z,x_3,b_B,b_3)\Big \},
\\
\non
F_{af_0}^{LL}&=&-\frac{8\pi C_Fm^4_Bf_B}{\sqrt{1-\eta}}\int^1_0 dx_3 dz \int_0^{1/\Lambda} bdb b_3db_3\phi_D(x_3,b_3)\non
     &&\times \Big \{ \big [\phi_0(z)r_{D}^2 (-2 (\eta -1) z-1)+(\eta -1) (z-1) \big ] \cdot E_a(t_e)h_e(z,x_3,b,b_3)S_t(z)\non
     &&+[(r_{D}^2-1) \phi_0(z) [ (\eta -1)^2 x_3-\eta  (\eta +r_{D}^2-1) ]\non
     &&-2 (\eta -1) r_{D} \phi_s(z) \sqrt{\eta -\eta  r_{D}^2}((r_{D}^2+(\eta-1)(1-x_3))]\non
     &&\cdot  E_a(t_f)h_f(z,x_3,b_3,b)S_t(|\eta(x_3-1)-x_3|)\Big \},
\end{eqnarray}
\begin{eqnarray}
M_{af_0}^{LL}&=&\frac{32\pi C_Fm^4_B}{\sqrt{6(\eta-\eta r_{D}^2)(1-\eta)}}\int^1_0 dx_B dzdx_3\int_0^{1/\Lambda} bdb b_Bdb_B\phi_B(x_B,b_B)\phi_D(x_3,b_3)\non
     &&\times \Big \{ [\phi_0(z) \sqrt{\eta -\eta  r_{D}^2} (r_{D}^2 ((\eta -2) (x_3(1-\eta)+ x_B-\eta (z-2)+1)\non
     &&+(\eta -1) (\eta  (x_3+z-1)-x_3-x_B))+(1-\eta) \eta r_{D} (1-r_{D}^2)((\phi_t(z)- \phi_s(z))\non
     &&\times ((1-x_3)(\eta -1)+x_B)-z((\phi_t(z)+ \phi_s(z)) \cdot E_n(t_g)h_g(x_B,z,x_3,b,b_B)\non
     &&+[\phi_0(z) (-\eta +r_{D}^2-1) \sqrt{\eta -\eta  r_{D}^2} (r_{D}^2 ((\eta -1)( x_3-z)x_B-1)\non
     &&+(\eta -1) (z-1))+(\eta -1) \eta r_{D}(1-r_{D}^2)(( \phi_t(z)- \phi_s(z)) (r_{D}^2-1)(z-1)\non
     &&+(( \phi_t(z)+\phi_s(z))(\eta +(1-\eta) x_3-x_B)))]\cdot E_n(t_h)h_h(x_B,z,x_3,b,b_B)\Big \},
\end{eqnarray}
with the ratio $r_D=m_{D^{(*)}}/m_{B_{(s)}}$ and the color factor $C_F=4/3$. $f_{D^{(*)}}$ ($f_B$) is the decay constant of the
$D^{(*)}$($B_{(s)}$) meson.

\item[$\bullet$]  $B_{(s)}\to \bar D(\phi\to)KK$
\begin{eqnarray}
F_{ef_2}^{LL}&=&\frac{8\pi C_Fm^4_Bf_D}{\sqrt{\eta-\eta r_D^2}}\int^1_0 dx_B dz \int_0^{1/\Lambda} bdb b_Bdb_B\phi_B(x_B,b_B)\non
     &&\times \Big \{[\phi_0(z) \sqrt{\eta -\eta  r_D^2} (r_D^2 (-2 \eta  (z+1)+2 z+1)+(\eta -1) (z+1))\non
     &&+\eta  (-r_D^2+1)( (\phi_t(z)+\phi_s(z))(1+2z(r_D^2 -1))(\eta -1)+r_D^2(\phi_t(z)\non
     &&-\phi_s(z)) )] \cdot E_e(t_a)h_a(x_B,z,b,b_B)S_t(z)+\sqrt{\eta -\eta  r_D^2} \big [\phi_0(z) (r_D^2 (\eta ^2-x_B)\non
     &&-(\eta -1) \eta )+2 \phi_s(z) \sqrt{\eta -\eta  r_D^2} (\eta +r_D^2 (-2 \eta+x_B+1)-1) \big ] \non
     && \cdot E_e(t_b)h_a(x_B,z,b_B,b)S_t(|x_B-\eta|) \Big \},
\\
\non
M_{ef_2}^{LL}&=&\frac{32\pi C_Fm^4_B}{\sqrt{6(\eta-\eta r_D^2)}}\int^1_0 dx_B dzdx_3\int_0^{1/\Lambda} b_3db_3 b_Bdb_B\phi_B(x_B,b_B)\phi_D(x_3,b_3)\non
     &&\times \Big  \{[\phi_0(z) (-\eta +r_D^2+1) \sqrt{\eta -\eta  r_D^2} ((1-r_D^2)((x_3+x_B-1)-(x_3+z)\eta) \non
     &&+\eta  (1-2r_D^2))+\eta  (-r_D^2+1)( \phi_t(z) ((\eta -1) z(1-r_D^2)-r_D^2 ((1-\eta)  x_3  \non
     &&+x_B))+ \phi_s(z) (r_D^2 ( x_3+2+z)-z)(\eta -1)] \cdot E_n(t_c)h_c(x_B,z,x_3,b_B,b_3)\non
     &&-\big [\phi_0(z) \sqrt{\eta -\eta  r_D^2} (-\eta +(2 \eta -1) r_D^2+1) ((r_D^2-1) z+(\eta -1) x_3+x_B)\non
     &&+\eta  (-r_D^2+1)(( \phi_t(z) -\phi_s(z))r_D^2 ((\eta -1) x_3+x_B)+(\eta -1) z(r_D^2-1) ( \phi_t(z)+\phi_s(z))) \big ] \non
     && \cdot E_n(t_d)h_d(x_B,z,x_3,b_B,b_3)\Big \},
\end{eqnarray}
\begin{eqnarray}
F_{af_2}^{LL}&=&\frac{8\pi C_Fm^4_Bf_B}{\sqrt{\eta-\eta r_D^2}}\int^1_0 dx_3 dz \int_0^{1/\Lambda} bdb b_3db_3\phi_D(x_3,b_3)\non
     &&\times \Big \{ \big [\phi_0(z) \sqrt{\eta -\eta  r_D^2} (r_D^2 (2 \eta -2 (\eta -1) z-1)+(\eta -1) (z-1))-2 \eta  r_D (r_D^2\non
     &&-1)^2 z (\phi_t(z)-\phi_s(z))+4 \eta  r_D(r_D^2-1)\phi_s(z) \big ] \cdot E_a(t_e)h_e(z,x_3,b,b_3) S_t(z)\non
     &&+\sqrt{\eta -\eta  r_D^2} \big  [\phi_0(z) (\eta  (\eta -\eta  r_D^2-1)+(\eta -1)^2 (r_D^2-1) x_3)+2 r_D \phi_s(z) \non
     && (\eta -r_D^2-\eta  x_3+x_3+1) \big ] \cdot E_a(t_f)h_f(z,x_3,b_3,b) S_t(|\eta(x_3-1)-x_3|)\Big \},  \quad
\end{eqnarray}
\begin{eqnarray}
M_{af_2}^{LL}&=&\frac{32\pi C_Fm^4_B}{\sqrt{6(\eta-\eta r_D^2)}}\int^1_0 dx_B dzdx_3\int_0^{1/\Lambda} bdb b_Bdb_B\phi_B(x_B,b_B)\phi_D(x_3,b_3)\non
     &&\times \Big \{ \big [\phi_0(z) \sqrt{\eta -\eta  r_D^2} (r_D^2 (\eta  (-\eta  (x_3+z-2)+x_3+x_B)-1)+(\eta -1) (\eta  (x_3\non
     &&+z-1)-x_3-x_B))+\eta  r_D ((\phi_t(z)-\phi_s(z) ) ((1-\eta)(x_3-1) +x_B) \non
     &&+(\phi_t(z)+\phi_s(z) )z-4\phi_s(z)) \big ]E_n(t_g)h_g(x_B,z,x_3,b,b_B)\non
     &&-\big [\phi_0(z) (-\eta +r_D^2-1) \sqrt{\eta -\eta  r_D^2} (r_D^2 (\eta  (x_3+z-2)-x_3+x_B-z+1)\non
     &&+( \eta-1) (1- z))+\eta  r_D(1-r_D^2) ((\phi_t(z)-\phi_s(z) )  (r_D^2-1)(z-1) +(\phi_t(z) \non
     &&+\phi_s(z) )(\eta-1) x_3+x_B-\eta) \big ] \cdot E_n(t_h)h_h(x_B,z,x_3,b,b_B)\Big \}.
\end{eqnarray}

\item[$\bullet$] $B_{(s)}\to \bar D^*(\phi\to)KK$

The formulas for the longitudinal component amplitudes $A_L$ are as follows:
\begin{eqnarray}
F_{a\phi,L}^{LL}&=&-\frac{8\pi C_Fm^4_Bf_B}{\sqrt{1-\eta}}\int^1_0 dx_3 dz \int_0^{1/\Lambda} bdb b_3db_3\phi_D(x_3,b_3)\non
     &&\times \Big \{ \left [\phi_0(z)r_{D}^2 (-2 (\eta -1) z-1)+(\eta -1) (z-1) \right ] \cdot E_a(t_e)h_e(z,x_3,b,b_3)S_t(z)\non
     &&+\big [(r_{D}^2-1) \phi_0(z) ((\eta -1)^2 x_3-\eta  (\eta +r_{D}^2-1))\non
     &&-2 (\eta -1) r_{D} \phi_s(z) \sqrt{\eta -\eta  r_{D}^2}((r_{D}^2+(\eta-1)(1-x_3)) \big ]\non
     &&\cdot  E_a(t_f)h_f(z,x_3,b_3,b)S_t(|\eta(x_3-1)-x_3|)\Big \},
\end{eqnarray}
\begin{eqnarray}
F_{e\phi,L}^{LL}&=&-\frac{8\pi C_Fm^4_Bf_D}{\sqrt{(\eta -\eta  r_{D}^2)(1-\eta)}}\int^1_0 dx_B dz \int_0^{1/\Lambda} bdb b_Bdb_B\phi_B(x_B,b_B)\non
     &&\times \Big \{ \big [\phi_0(z) \sqrt{\eta -\eta  r_{D}^2} (r_{D}^2 (1-2 (\eta -1) z)+(\eta -1) (z+1))\non
     &&+\eta  (-r_{D}^2+1)( (\phi_s(z)+\phi_t(z))(1+2z(r_{D}^2 -1))(\eta -1)+r_{D}^2(-\phi_t(z)+\phi_s(z))) \big ]\non
     && \cdot  E_e(t_a)h_a(x_B,z,b,b_B)S_t(z)\non
     &&- \sqrt{\eta -\eta  r_{D}^2} \big [\phi_0(z) (-(r_{D}^2 ((\eta -2) \eta +x_B)-(\eta -1) \eta ))\non
     &&+2 \phi_s(z) \sqrt{\eta -\eta  r_{D}^2} (-\eta +r_{D}^2 (x_B-1)+1) \big ]\non
     &&\cdot  E_e(t_b)h_a(x_B,z,b_B,b)S_t(|x_B-\eta|) \Big \},
\end{eqnarray}
\begin{eqnarray}
M_{e\phi,L}^{LL}&=&\frac{32\pi C_Fm^4_B}{\sqrt{6(\eta-\eta r_{D}^2)(1-\eta)}}\int^1_0 dx_B dzdx_3\int_0^{1/\Lambda} b_3db_3 b_Bdb_B\phi_B(x_B,b_B)\phi_D(x_3,b_3)\non
     &&\times \Big \{ \big [\phi_0(z) (\eta +r_{D}^2-1) \sqrt{\eta -\eta  r_{D}^2} ((\eta(1-x_3-z) +r_{D}^2 \eta  (x_3+z-2)\non
     &&+(1-r_{D}^2)(x_3+x_B-1))+\eta  (-r_{D}^2+1)( \phi_s(z) ((\eta -1) z(1-r_{D}^2)-r_{D}^2 ((1 \non
     &&-\eta)  x_3 +x_B))+ \phi_t(z) (r_{D}^2 ( x_3+2+z)-z)(\eta -1) \big ] \cdot E_n(t_c)h_c(x_B,z,x_3,b_B,b_3)\non
     &&-\big [\phi_0(z) (\eta +r_{D}^2-1) \sqrt{\eta -\eta  r_{D}^2} ((r_{D}^2-1) z+(\eta -1) x_3+x_B)\non
     &&+\eta  (-r_{D}^2+1)(( \phi_t(z)-\phi_s(z)) r_{D}^2 ((\eta -1) x_3+x_B)+( \phi_t(z)\non
     && +\phi_s(z))z(1-\eta)(r_{D}^2-1) \big ] \cdot E_n(t_d)h_d(x_B,z,x_3,b_B,b_3)\Big \},
\end{eqnarray}
\begin{eqnarray}
M_{a\phi,L}^{LL}&=&\frac{32\pi C_Fm^4_B}{\sqrt{6(\eta-\eta r_{D}^2)(1-\eta)}}\int^1_0 dx_B dzdx_3\int_0^{1/\Lambda} bdb b_Bdb_B\phi_B(x_B,b_B)\phi_D(x_3,b_3)\non
     &&\times \Big \{ \Big [\phi_0(z) \sqrt{\eta -\eta  r_{D}^2}\Big [  r_{D}^2  ((\eta -2) (x_3(1-\eta)+ x_B-\eta (z-2)+1)\non
     &&+(\eta -1) (\eta  (x_3+z-1)-x_3-x_B))+(1-\eta) \eta r_{D} [ (\phi_t(z)- \phi_s(z)) [  (1\non
     &&-x_3)(\eta -1)+x_B ] - z (\phi_t(z)- \phi_s(z)  ) ]  \Big ] \Big ]\cdot E_n(t_g)h_g(x_B,z,x_3,b,b_B)\non
     &&+\Big [\phi_0(z) (-\eta +r_{D}^2-1) \sqrt{\eta -\eta  r_{D}^2} (r_{D}^2 ((\eta -1)( x_3-z)x_B-1)\non
     &&+(\eta -1) (z-1))+(\eta -1) \eta r_{D}(1-r_{D}^2)(( \phi_t(z)- \phi_s(z)) (r_{D}^2-1)(z-1)\non
     &&+(( \phi_t(z)+ \phi_s(z))(\eta +(1-\eta) x_3-x_B))) \Big ] \cdot E_n(t_h)h_h(x_B,z,x_3,b,b_B) \Big \}.
\end{eqnarray}

The transverse polarization amplitudes $A_{N,T}$ are of the following form:
\begin{eqnarray}
F_{e\phi,N}^{LL}&=&-8\pi C_Fm^4_Bf_{D^*}r_{D}\int^1_0 dx_B dz \int_0^{1/\Lambda} bdb b_Bdb_B\phi_B(x_B,b_B)\non
     &&\times \Big \{[\sqrt{\eta -\eta  r_{D}^2} (r_{D}^2 (-z)+z+2) \phi_a(z)+\phi_T(z) (\eta +(r_{D}^2-1) (2 \eta  z-1))\non
     &&+(r_{D}^2-1) z \sqrt{\eta -\eta  r_{D}^2} \phi_v(z)] \cdot E_e(t_a)h_a(x_B,z,b,b_B)S_t(z)\non
     &&-\sqrt{\eta -\eta  r_{D}^2} \big [\phi_a(z) (-\eta +r_{D}^2+x_B-1)+\phi_v(z)(\eta +r_{D}^2-x_B-1) \big ] \non
     && \cdot E_e(t_b)h_a(x_B,z,b_B,b) S_t(|x_B-\eta| ) \Big \},
\\
\non
F_{e\phi,T}^{LL}&=&-8\pi C_Fm^4_Bf_{D^*}r_{D}\int^1_0 dx_B dz \int_0^{1/\Lambda} bdb b_Bdb_B\phi_B(x_B,b_B)\non
     &&\times \Big \{ \big [(-r_{D}^2+1) z \sqrt{\eta -\eta  r_{D}^2} \phi_a(z)+\phi_T(z) (\eta +(2 \eta z+1)(r_{D}^2-1))\non
     &&-\sqrt{\eta -\eta  r_{D}^2} (r_{D}^2 (-z)+z+2) \phi_v(z) \big ] \cdot E_e(t_a)h_a(x_B,z,b,b_B)S_t(z)\non
     &&+\sqrt{\eta -\eta  r_{D}^2} \big [(\phi_a(z)+\phi_v(z)) (1-r_{D}^2)+(\phi_v(z)-\phi_a(z))(\eta-x_B) \big ]\non
     &&\cdot  E_e(t_b)h_a(x_B,z,b_B,b)S_t(|x_B-\eta|) \Big \},
\end{eqnarray}
\begin{eqnarray}
M_{e\phi,N}^{LL}&=&16\sqrt{\frac{2}{3}}\pi C_Fm^4_Br_{D}\int^1_0 dx_B dzdx_3\int_0^{1/\Lambda} b_3db_3 b_Bdb_B\phi_B(x_B,b_B)\phi_D(x_3,b_3)\non
     &&\times \Big \{ \left [\phi_T(z)(\eta +r_{D}^2 \eta  (x_3+z-2)+(1-r_{D}^2 )(x_3+x_B-1)+\eta (1-x_3-z)) \right ]\non
     &&\cdot E_n(t_c)h_c(x_B,z,x_3,b_B,b_3)\non
     &&- \Big [ (2 \sqrt{\eta -\eta  r_{D}^2} \phi_a(z) ((r_{D}^2-1) z+(\eta -1) x_3+x_B)+(r_{D}^2-1) \phi_T(z) ((\eta \non
     &&-1) x_3+x_B-\eta  z)) \Big ]\cdot E_n(t_d)h_d(x_B,z,x_3,b_B,b_3) \Big \},
\\
\non
M_{e\phi,T}^{LL}&=&16\sqrt{\frac{2}{3}}\pi C_Fm^4_Br_{D}\int^1_0 dx_B dzdx_3\int_0^{1/\Lambda} b_3db_3 b_Bdb_B\phi_B(x_B,b_B)\phi_D(x_3,b_3)\non
     &&\times \Big \{ \left [-\phi_T(z)((x_3+x_B-1)(1-r_{D}^2)+\eta ((r_{D}^2-1) (x_3 -z)+1)) \right ]\non
     && \cdot E_n(t_c)h_c(x_B,z,x_3,b_B,b_3)\non
     &&-\big [-((r_{D}^2-1) \phi_T(z) (\eta  (x_3+z)-x_3+x_B)+2 \sqrt{\eta -\eta  r_{D}^2} \phi_v(z)((r_{D}^2-1)\non
     &&z+(\eta -1) x_3+x_B))\big ]\cdot E_n(t_d)h_d(x_B,z,x_3,b_B,b_3) \Big \},
\\
\non
F_{a\phi,N}^{LL}&=&8\pi C_Fm^4_Bf_Br_{D}\sqrt{\eta -\eta  r_{D}^2}\int^1_0 dx_3 dz \int_0^{1/\Lambda} bdb b_3db_3\phi_D(x_3,b_3)\non
     &&\times \Big \{ \left [ (r_{D}^2-1) z (\phi_a(z)-\phi_v(z))+2 \phi_a(z) \right ]\cdot E_a(t_e)h_e(z,x_3,b,b_3)S_t(z)\non
     &&+\left [\phi_a(z) (r_{D}^2+\eta  (x_3-1)-x_3-1)-\phi_v(z)(r_{D}^2+(1-\eta)(x_3-1)) \right ]\non
     &&\cdot  E_a(t_f)h_f(z,x_3,b_3,b)S_t(|\eta(x_3-1)-x_3|)\Big \},
\\
\non
F_{a\phi,T}^{LL}&=&8\pi C_Fm^4_Bf_Br_{D}\sqrt{\eta -\eta  r_{D}^2}\int^1_0 dx_3 dz \int_0^{1/\Lambda} bdb b_3db_3\phi_D(x_3,b_3)\non
     &&\times \Big \{ \left [(r_{D}^2-1) z (\phi_a(z)-\phi_v(z))-2 \phi_v(z) \right ] \cdot E_a(t_e)h_e(z,x_3,b,b_3)S_t(z)\non
     &&+\left [ \phi_a(z) (r_{D}^2+(1-\eta)(x_3-1)+\phi_v(z)(\eta -r_{D}^2-\eta  x_3+x_3+1)) \right ]\non
     &&\cdot E_a(t_f)h_f(z,x_3,b_3,b)S_t(|\eta(x_3-1)-x_3|)\Big \},
\\
\non
M_{a\phi,N}^{LL}&=&16\sqrt{\frac{2}{3}}\pi C_Fm^4_B\int^1_0 dx_B dzdx_3\int_0^{1/\Lambda} bdb b_Bdb_B\phi_B(x_B,b_B)\phi_D(x_3,b_3)\non
     &&\times \Big \{ \big [2 r_{D} \sqrt{\eta -\eta  r_{D}^2} \phi_a(z)+\phi_T(z) (-r_{D}^2 ((1-\eta)(x_3+z\eta)+x_B+(\eta)^2-1) \non
     &&-(\eta -1) \eta  z)+2 r_{D} \sqrt{\eta -\eta  r_{D}^2} \phi_v(z)(r_{D}^2 (z-1)\non
     &&+\eta  (x_3-1)-x_3-x_B-z+1) \big ]\cdot E_n(t_g)h_g(x_B,z,x_3,b,b_B)\non
     &&+ \big [ (r_{D}^2-1) \phi_T(z) (r_{D}^2 (\eta  (x_3-1)-x_3+x_B)+\eta(1-\eta)(z-1)\non
     &&+2 r_{D} \sqrt{\eta -\eta  r_{D}^2} \phi_v(z)((r_{D}^2-1) (z-1)+\eta(x_3-1)-x_3+x_B) )  \big ]\non
     && \cdot  E_n(t_h)h_h(x_B,z,x_3,b,b_B)\Big \},
\end{eqnarray}
\begin{eqnarray}
M_{a\phi,T}^{LL}&=&16\sqrt{\frac{2}{3}}\pi C_Fm^4_B\int^1_0 dx_B dzdx_3\int_0^{1/\Lambda} bdb b_Bdb_B\phi_B(x_B,b_B)\phi_D(x_3,b_3)\non
     &&\times \Big \{  \big [\phi_T(z) (-(r_{D}^2 (x_B+(\eta -1) ((\eta  (z-1)+1-x_3))-(\eta -1) \eta  z))\non
     &&-2 r_{D} \sqrt{\eta -\eta  r_{D}^2} \phi_v(z)(\eta +r_{D}^2 (z-1)-\eta  x_3+x_3+x_B-z+3) \big ]\non
     &&\cdot  E_n(t_g)h_g(x_B,z,x_3,b,b_B)\non
     &&+\big [(r_{D}^2-1) \phi_T(z) (r_{D}^2 (\eta  (x_3-1)-x_3+x_B)+(\eta -1) \eta  (z-1))\non
     &&+2 r_{D} \sqrt{\eta -\eta  r_{D}^2} \phi_v(z)(\eta +r_{D}^2 (z-1)-\eta  x_3+x_3-x_B-z+1) \big ]\non
     && \cdot  E_n(t_h)h_h(x_B,z,x_3,b,b_B) \Big \}.
\end{eqnarray}
\end{enumerate}

For the decays involved $ D^{(*)}$ mesons, similarly, the corresponding decay amplitudes can be obtained  by evaluating the Feyman diagrams in Fig.~\ref{fig:fig2}.
\begin{enumerate}
\item[$\bullet$] $B_{(s)}\to D(f_0\to)KK$
\begin{eqnarray}
F_{ef_0}^{LL}&=&\ 8\pi C_Fm^4_Bf_D\int^1_0 dx_B dz \int_0^{1/\Lambda} bdb b_Bdb_B\phi_B(x_B,b_B)\non
     &&\times \Big \{ \big [\phi_0(z)  (r_D^2 (-2 \eta  (z+1)+2 z+1)+(\eta -1) (z+1))\non
     &&+\sqrt{\eta -\eta  r_D^2}((\phi_t(z)+\phi_s(z)) (\eta -1)(1+2z(r_D^2-1))+r_D^2(\phi_t(z)\non
     &&-\phi_s(z))) \big ]\cdot  E_e(t_a)h_a(x_B,z,b,b_B)S_t(z)+ \big [\phi_0(z) (r_D^2 (\eta ^2-x_B)\non
     &&-(\eta -1) \eta )+2 \phi_s(z)\sqrt{\eta -\eta  r_D^2}(\eta +r_D^2 (-2 \eta +x_B+1)-1) \big ]\non
     &&\cdot E_e(t_b)h_a(x_B,z,b_B,b)S_t(|x_B-\eta|) \Big \},
\\
\non
M_{ef_0}^{LL}&=&\frac{32\pi C_Fm^4_B}{\sqrt{6}}\int^1_0 dx_B dzdx_3\int_0^{1/\Lambda} b_3db_3 b_Bdb_B\phi_B(x_B,b_B)\phi_D(x_3,b_3)\non
     &&\times \Big \{ \big [\phi _0(z)((1-\eta ) x_3(\eta (r_D^2-1)+1)-\eta (r_D^2-1)(z(-\eta +r_D^2+1)\non
     &&+x_B)-x_B)+\sqrt{\eta -\eta  r_D^2}(r_D^2 ((\eta -1) x_3+ x_B )(\phi _s(z)+\phi _t(z))\non
     &&-(\eta -1) z(r_D^2 -1)(\phi _s(z)-\phi _t(z))) \big ]\cdot E_n(t_c)h_c(x_B,z,x_3,b_B,b_3)\non
     && +\big [\phi _0(z) (-\eta +(2 \eta -1) r_D^2+1)(\eta +r_D^2 (z-1)-\eta  x_3+x_3+x_B-z-1)\non
     &&+\sqrt{\eta -\eta  r_D^2}(r_D^2 (\phi _t(z) +\phi _s(z) )((-\eta  x_3+x_3+x_B)-2\phi _s(z) (\eta -1))
     \non
     &&+(r_D^2 -1)(\eta -1)z (\phi _t(z)-\phi _s(z)) \big ]  \cdot E_n(t_d)h_d(x_B,z,x_3,b_B,b_3) \Big \},
\end{eqnarray}
\begin{eqnarray}
F_{aD}^{LL}&=& -8\pi C_Fm^4_Bf_B\int^1_0 dx_3 dz \int_0^{1/\Lambda} bdb b_3db_3\phi_D(x_3,b_3)\non
     &&\times \Big \{ \big [\phi _0(z)(\eta (\eta -\eta  r_D^2-1)+(\eta -1)^2(r_D^2-1) x_3)\non
     &&+2 r_D \phi _s(z) \sqrt{\eta -\eta  r_D^2}(-\eta +r_D^2+(\eta -1) x_3-1) \big ] \cdot E_a(t_e)h_e(z,x_3,b,b_3)S_t(z)\non
     && -\big [ (\eta -1) \phi _0(z)(r_D^2 (\eta -2 z+1)+z)\non
     &&+2 r_D \sqrt{\eta -\eta  r_D^2} ((\phi _t(z) +\phi _s(z) )(\eta -1)+z(\phi _t(z)\non
     &&-\phi _s(z) )) \big ] \cdot E_a(t_f)h_f(z,x_3,b_3,b)S_t(|\eta(x_3-1)-x_3|) \Big\},
\\
\non
M_{aD}^{LL}&=&\frac{32\pi C_Fm^4_B}{\sqrt{6}}\int^1_0 dx_B dzdx_3\int_0^{1/\Lambda} bdb b_Bdb_B\phi_B(x_B,b_B)\phi_D(x_3,b_3)\non
     &&\times \Big \{ \big [\phi _0(z)(r_D^2((1-\eta ^2) x_3+(\eta -1) x_B+(\eta ^2+\eta -2) z-1)\non
     &&-(\eta -1) ((\eta  +1)(x_B+z)-\eta  ))-r_D\sqrt{\eta -\eta  r_D^2}((\phi _t(z)+\phi _s(z)) (\eta \non
     && -1) x_3+(\phi _t(z)-\phi _s(z))(x_B+z)-2\phi _s(z) ) \big ] \cdot E_n(t_g)h_g(x_B,z,x_3,b,b_B)\non
     &&-\big [\phi _0(z)(\eta (-\eta +r_D^2+1)(-((r_D^2-1) z+x_B))-(\eta -1) x_3(\eta \non
     &&\times (r_D^2-1)+1))-r_D \sqrt{\eta -\eta  r_D^2}((\phi _t(z) +\phi _s(z))((r_D^2-1) z+x_B)\non
     &&+( \phi _s(z) -\phi _t(z)) (\eta  -1)x_3) \big ] \cdot E_n(t_h)h_h(x_B,z,x_3,b,b_B) \Big \}.
\end{eqnarray}
\item[$\bullet$] $B_{(s)}\to D^*(f_0\to)KK$
\begin{eqnarray}
F_{ef_0}^{LL}&=&-\frac{8\pi C_Fm^4_Bf_{D^*}}{\sqrt{(1-\eta)}}\int^1_0 dx_B dz \int_0^{1/\Lambda} bdb b_Bdb_B\phi_B(x_B,b_B)\non
     &&\times \Big \{ \big [\phi_0(z)(r_{D}^2 (1-2 (\eta -1) z)+(\eta -1) (z+1))+\sqrt{\eta -\eta  r_{D}^2}  \non
     &&\times((\phi_t(z)+\phi_s(z)) (\eta -1) (1+2z(r_{D}^2 -1))+r_{D}^2(\phi_s(z)\non
     &&-\phi_t(z))) \big ] \cdot E_e(t_a)h_a(x_B,z,b,b_B)S_t(z)- \big [\phi_0(z) (-(r_{D}^2 ((\eta -2) \eta +x_B)\non
     &&-(\eta -1) \eta ))+2 \phi_s(z)\sqrt{\eta -\eta  r_{D}^2}(-\eta +r_{D}^2 (x_B-1)+1) \big ] \non
     &&\cdot  E_e(t_b)h_a(x_B,z,b_B,b)S_t(|x_B-\eta|) \Big \},
\\
\non
M_{ef_0}^{LL}&=&-\frac{32\pi C_Fm^4_B}{\sqrt{6(1-\eta)}}\int^1_0 dx_B dzdx_3\int_0^{1/\Lambda} b_3db_3 b_Bdb_B\phi_B(x_B,b_B)\phi_D(x_3,b_3)\non
     &&\times \Big \{ \big [(r_{D}^2-1) \phi _0(z)(\eta +r_{D}^2-1) ((\eta -1) x_3+x_B-\eta  z)\non
     &&+\sqrt{\eta -\eta  r_{D}^2}((\phi _t(z)+\phi _s(z) ) r_{D}^2 ((\eta -1) x_3+x_B)\non
     &&+(\phi _t(z)-\phi _s(z) )(1-r_{D}^2)(\eta  -1)z) \big ] \cdot E_n(t_c)h_c(x_B,z,x_3,b_B,b_3)\non
     &&+\big [\phi_0(z) (\eta +r_{D}^2-1)(\eta +r_{D}^2 (z-1)-(\eta-1) x_3\non
     &&+x_B-z-1)+\sqrt{\eta -\eta  r_{D}^2}( (\phi_t(z)+\phi_s(z))((\eta -1)(1-r_{D}^2) z\non
     &&+(\phi_t(z)-\phi_s(z))r_{D}^2 (-(\eta -1) x_3+x_B)\non
     &&+2r_{D}^2\phi_t(z)(\eta  -1)) \big ] \cdot E_n(t_d)h_d(x_B,z,x_3,b_B,b_3) \Big \},
\end{eqnarray}
\begin{eqnarray}
F_{aD}^{LL}&=&\frac{8\pi C_Fm^4_Bf_B}{\sqrt{1-\eta}}\int^1_0 dx_3 dz \int_0^{1/\Lambda} bdb b_3db_3\phi_D(x_3,b_3)\non
     &&\times \Big \{ \big [(r_{D}^2-1) \phi_0(z) ((\eta -1)^2 x_3-\eta  (\eta +r_{D}^2-1))+2 (\eta -1) r_{D} \phi_s(z)\non
     &&\times\sqrt{\eta -\eta  r_{D}^2} (\eta +r_{D}^2-\eta x_3+x_3-1) \big ] \cdot E_a(t_e)h_e(z,x_3,b,b_3)S_t(z)\non
     &&+(1- \eta ) \big [\phi_0(z) (z-r_{D}^2 (\eta +2 z-1)) \big ] \non
     && \cdot  E_a(t_f)h_f(z,x_3,b_3,b)S_t(|\eta(x_3-1)-x_3|) \Big \},
\\
\non
M_{aD}^{LL}&=&-\frac{32\pi C_Fm^4_B}{\sqrt{6(1-\eta)}}\int^1_0 dx_B dzdx_3\int_0^{1/\Lambda} bdb b_Bdb_B\phi_B(x_B,b_B)\phi_D(x_3,b_3)\non
     &&\times \Big \{ \big [\phi _0(z)(r_{D}^2(\eta ^2 (x_3+z)-x_3+(\eta -1) x_B+(\eta -2) z+1)-(\eta ^2-1) \non
     &&\times (x_B+z)+\eta (\eta -1))+ \sqrt{\eta -\eta  r_{D}^2}  (\eta-1)(-r_{D}) ( (\phi_t(z)-\phi_s(z)) (1\non
     &&-\eta) x_3-2\phi_t(z)+(\phi_t(z)+\phi_s(z))(x_B+z)) \big ]\cdot E_n(t_g)h_g(x_B,z,x_3,b,b_B)\non
     &&+\big [\phi_0(z) (\eta +r_{D}^2-1)((\eta -1) (r_{D}^2-1) x_3-\eta  ((r_{D}^2-1) z+x_B))\non
     &&+(\eta -1) (-r) \sqrt{\eta -\eta  r_{D}^2}((\phi_t(z)-\phi_s(z)) ((r_{D}^2-1) z+x_B)+(\phi_t(z)\non
     &&+\phi_s(z))((\eta-1)x_3)) \big] \cdot E_n(t_h)h_h(x_B,z,x_3,b,b_B) \Big \}.
\end{eqnarray}
\item[$\bullet$] $B_{(s)}\to D(\phi\to)KK$
\begin{eqnarray}
F_{e\phi}^{LL}&=&\frac{8\pi C_Fm^4_Bf_D}{\sqrt{\eta-\eta r_D^2}}\int^1_0 dx_B dz \int_0^{1/\Lambda} bdb b_Bdb_B\phi_B(x_B,b_B)\non
     &&\times \Big \{ \big [\phi_0(z) \sqrt{\eta -\eta  r_D^2} (r_D^2 (-2 \eta  (z+1)+2 z+1)+(\eta -1) (z+1))\non
     &&+\eta  (-(r_D^2-1))((\phi_t(z)+\phi_s(z)) (\eta -1)(1+2z(r_D^2-1))\non
     &&+r_D^2(\phi_t(z)-\phi_s(z))) \big ] \cdot E_e(t_a)h_a(x_B,z,b,b_B)S_t(z)+\sqrt{\eta-\eta r_D^2} \non
     &&\times \big [\phi_0(z) (r_D^2(\eta ^2-x_B)-(\eta -1) \eta )+2 \phi_s(z)\sqrt{\eta -\eta  r_D^2}(\eta +r_D^2 (-2 \eta \non
     &&\times+x_B+1)-1) \big ] \cdot E_e(t_b)h_a(x_B,z,b_B,b)S_t(|x_B-\eta|) \Big \},
\\
\non
M_{e\phi}^{LL}&=&\frac{32\pi C_Fm^4_B}{\sqrt{6(\eta-\eta r_D^2)}}\int^1_0 dx_B dzdx_3\int_0^{1/\Lambda} b_3db_3 b_Bdb_B\phi_B(x_B,b_B)\phi_D(x_3,b_3)\non
     &&\times \Big \{ (r_D^2-1) \big [\phi_0(z) (-(-\eta +r_D^2+1)) \sqrt{\eta -\eta  r_D^2} ((\eta -1) x_3+x_B-\eta  z)\non
     &&+\eta  (\phi_t(z)+\phi_s(z)) (r_D^2 ((\eta -1) x_3+x_B ))+ (\phi_t(z)-\phi_s(z))\non
     &&\times(r_D^2 -1)(\eta -1)z \big ] \cdot E_n(t_c)h_c(x_B,z,x_3,b_B,b_3)\non
     && -\big [ (\phi_0(z) \sqrt{\eta -\eta  r_D^2} (-\eta +(2 \eta -1) r_D^2+1)((\eta-1)(1-x_3) +r_D^2 (z-1)\non
     &&+x_B-z)+\eta  (r_D^2-1)( (\phi_t(z)+\phi_s(z))(\eta -1) z(1-r_D^2) \non
     &&-(\phi_t(z)-\phi_s(z))r_D^2 (-\eta  x_3+x_3+x_B)+2r_D^2\phi_s(z)(\eta  -1)) \big ]\non
     &&\cdot E_n(t_d)h_d(x_B,z,x_3,b_B,b_3)\Big \},
\end{eqnarray}
\begin{eqnarray}
F_{aD}^{LL}&=&-\frac{8\pi C_Fm^4_Bf_B}{\sqrt{\eta-\eta r_D^2}}\int^1_0 dx_3 dz \int_0^{1/\Lambda} bdb b_3db_3\phi_D(x_3,b_3)\non
     &&\times \Big  \{ \big [\phi_0(z) \sqrt{\eta -\eta  r_D^2} (\eta  (\eta -\eta r_D^2-1)+(\eta -1)^2(r_D^2-1) x_3)  \non
     &&-2 \eta r_D (r_D^2-1)\phi_s(z)(-\eta+r_D^2+(\eta -1) x_3-1) \big ] \cdot E_a(t_e)  h_e(z,x_3,b,b_3)S_t(z)\non
     && -\big [ (\eta -1) \phi_0(z)\sqrt{\eta -\eta  r_D^2} \left ( r_D^2 (\eta -2 z+1)+z \right )\non
     && +2 \eta r_D (\phi_t(z)+\phi_s(z))(\eta -1)+2 \eta  r_D z (\phi_t(z) -\phi_s(z)) \big ]\non
     && \cdot  E_a(t_f)h_f(z,x_3,b_3,b)S_t(|\eta(x_3-1)-x_3|) \Big \},
\\
\non
M_{aD}^{LL}&=&\frac{32\pi C_Fm^4_B}{\sqrt{6(\eta-\eta r_D^2)}}\int^1_0 dx_B dzdx_3\int_0^{1/\Lambda} bdb b_Bdb_B\phi_B(x_B,b_B)\phi_D(x_3,b_3)\non
     &&\times \Big \{ \big [\phi_0(z) \sqrt{\eta -\eta  r_D^2} (r_D^2 (\eta ^2 (-x_3)+x_3+(\eta -1) x_B+(\eta ^2+\eta -2) z-1)\non
     &&-(\eta -1) (\eta  (x_B+z-1)+x_B+z))+\eta r_D(-( \phi_t(z)+\phi_s(z))((\eta -1) x_3)\non
     &&+(\phi_s(z)-\phi_t(z))(x_B+z)+2\phi_s(z) \big ] \cdot E_n(t_g)h_g(x_B,z,x_3,b,b_B)\non
     &&-\big [\phi_0(z) (-\eta +r_D^2+1) \sqrt{\eta -\eta  r_D^2} ((\eta -1) (r_D^2-1) x_3-\eta  ((r_D^2-1) z+x_B))\non
     &&+\eta  r (r_D^2-1)((\phi_t(z)+\phi_s(z)) ((r_D^2-1) z+x_B)+(\eta -1)x_3\times (\phi_s(z)-\phi_t(z))) \big ]\non
     && \cdot E_n(t_h)h_h(x_B,z,x_3,b,b_B) \Big \}.
\end{eqnarray}
\\
\item[$\bullet$] $B_{(s)}\to D^*(\phi\to)KK$

The formulas for the longitudinal amplitudes $A_L$ are as follows:
\begin{eqnarray}
F_{ef_2,L}^{LL}&=&-\frac{8\pi C_Fm^4_Bf_{D^*}}{\sqrt{(\eta -\eta  r_{D}^2)(1-\eta)}}\int^1_0 dx_B dz \int_0^{1/\Lambda} bdb b_Bdb_B\phi_B(x_B,b_B)\non
     &&\times \{[\phi_0(z) \sqrt{\eta -\eta  r_{D}^2} (r_{D}^2 (1-2 (\eta -1) z)+(\eta -1) (z+1))+\eta  (-(r_{D}^2-1)) \non
     &&\times((\phi_t(z)+\phi_s(z)) (\eta -1) (1+2z(r_{D}^2 -1)+r_{D}^2(\phi_s(z)-\phi_t(z)))]\non
     &&E_e(t_a)h_a(x_B,z,b,b_B)S_t(z)- \sqrt{\eta -\eta  r_{D}^2}[\phi_0(z) (-(r_{D}^2 ((\eta -2) \eta +x_B)\non
     &&-(\eta -1) \eta ))+2 \phi_s(z)\sqrt{\eta -\eta  r_{D}^2}(-\eta +r_{D}^2 (x_B-1)+1)] \non
     &&\times E_e(t_b)h_a(x_B,z,b_B,b)S_t(|x_B-\eta|)\},
\\
\non
F_{aD,L}^{LL}&=&\frac{8\pi C_Fm^4_Bf_B}{\sqrt{1-\eta}}\int^1_0 dx_3 dz \int_0^{1/\Lambda} bdb b_3db_3\phi_D(x_3,b_3)\non
     &&\times \{[(r_{D}^2-1) \phi_0(z) ((\eta -1)^2 x_3-\eta  (\eta +r_{D}^2-1))+2 (\eta -1) r_{D} \phi_s(z)\non
     &&\times\sqrt{\eta -\eta  r_{D}^2} (\eta +r_{D}^2-\eta x_3+x_3-1)]E_a(t_e)h_e(z,x_3,b,b_3)S_t(z)\non
     &&+(1- \eta )[\phi_0(z) (z-r_{D}^2 (\eta +2 z-1))]E_a(t_f)h_f(z,x_3,b_3,b)\non
     &&S_t(|\eta(x_3-1)-x_3|)\},
\end{eqnarray}
\begin{eqnarray}
M_{ef_2,L}^{LL}&=&-\frac{32\pi C_Fm^4_B}{\sqrt{6(\eta-\eta r_{D}^2)(1-\eta)}}\int^1_0 dx_B dzdx_3\int_0^{1/\Lambda} b_3db_3 b_Bdb_B\phi_B(x_B,b_B)\phi_D(x_3,b_3)\non
     &&\times \{[\phi_0(z) (\eta +r_{D}^2-1) \sqrt{\eta -\eta  r_{D}^2} ((\eta -1) x_3+x_B-\eta  z)\non
     &&-\eta ( (\phi_t(z)+\phi_s(z)) (r_{D}^2 ((\eta -1) x_3+x_B ))+ (\phi_s(z)-\phi_t(z))\non
     &&\times(r_{D}^2 -1)(\eta -1)z)]E_n(t_c)h_c(x_B,z,x_3,b_B,b_3)\non
     &&+[\phi_0(z) (\eta +r_{D}^2-1)\sqrt{\eta -\eta  r_{D}^2}(\eta +r_{D}^2 (z-1)-\eta  x_3+x_3\non
     &&+x_B-z-1)+\eta  (r_{D}^2-1)( (\phi_t(z)+\phi_s(z))((\eta -1) r_{D}^2 z\non
     &&- (\eta-1)z )-(\phi_t(z)-\phi_s(z))r_{D}^2 (-\eta  x_3+x_3+x_B)\non
     &&-2r_{D}^2\phi_t(z)(\eta  -1))]E_n(t_d)h_d(x_B,z,x_3,b_B,b_3)\},
\\
\non
M_{aD,L}^{LL}&=&-\frac{32\pi C_Fm^4_B}{\sqrt{6(\eta-\eta r_{D}^2)(1-\eta)}}\int^1_0 dx_B dzdx_3\int_0^{1/\Lambda} bdb b_Bdb_B\phi_B(x_B,b_B)\phi_D(x_3,b_3)\non
     &&\times \{ [\phi_0(z) \sqrt{\eta -\eta  r_{D}^2} (r_{D}^2 (\eta ^2 (x_3+z)-x_3+(\eta -1) x_B+(\eta -2) z+1)\non
     &&-(\eta -1) (\eta  (x_B+z-1)+x_B+z))+(\eta -1) \eta  r_{D}( (\phi_t(z)+\phi_s(z)) (1\non
     &&-\eta) x_3-2\phi_t(z)+(\phi_t(z)-\phi_s(z))(x_B+z))]E_n(t_g)h_g(x_B,z,x_3,b,b_B)\non
     &&+[\phi_0(z) (\eta +r_{D}^2-1) \sqrt{\eta -\eta  r_{D}^2} ((\eta -1) (r_{D}^2-1) x_3-\eta  ((r_{D}^2-1) z+x_B))\non
     &&+(\eta -1) \eta  (-r_{D}) (r_{D}^2-1)((\phi_t(z)+\phi_s(z)) ((r_{D}^2-1) z+x_B)+(\phi_t(z)\non
     &&-\phi_s(z))((\eta-1)x_3))]E_n(t_h)h_h(x_B,z,x_3,b,b_B)\}.
\end{eqnarray}
The expressions of the transverse component $A_{N,T}$ are given by:
\begin{eqnarray}
F_{ef_2,N}^{LL}&=&-8\pi C_Fm^4_Bf_{D^*}r_{D}\int^1_0 dx_B dz \int_0^{1/\Lambda} bdb b_Bdb_B\phi_B(x_B,b_B)\non
     &&\times \{[\sqrt{\eta -\eta  r_{D}^2} (((r_{D}^2 -1)z)(\phi _v(z)-\phi _a(z))+2 \phi _a(z))\non
     &&+\phi _T(z)(\eta +(r_{D}^2-1) (2 \eta  z-1)]E_e(t_a)h_a(x_B,z,b,b_B)S_t(z)\non
     &&-\sqrt{\eta -\eta  r_{D}^2}[(\phi _a(z)-\phi _v(z))(-\eta +x_B)
     +(\phi _a(z)+\phi_v(z))\non &&\times(r_{D}^2-1)]E_e(t_b)h_a(x_B,z,b_B,b)S_t(|x_B-\eta|)\},
\\
\non
F_{ef_2,T}^{LL}&=&-8\pi C_Fm^4_Bf_{D^*}r_{D}\int^1_0 dx_B dz \int_0^{1/\Lambda} bdb b_Bdb_B\phi_B(x_B,b_B)\non
     &&\times \{[-\sqrt{\eta -\eta  r_{D}^2}((r_{D}^2-1) z ( \phi _a(z)-\phi _v(z))+2\phi _v(z))\non
     &&+\phi _T(z)(\eta +(2 \eta z+1)(r_{D}^2-1))]E_e(t_a)h_a(x_B,z,b,b_B)S_t(z)\non
     &&-\sqrt{\eta -\eta  r_{D}^2}[(\phi _a(z)-\phi _v(z))(-\eta +x_B)-(\phi _v(z)+\phi _a(z))  \non
     &&\times(r_{D}^2-1)]E_e(t_b)h_a(x_B,z,b_B,b)S_t(|x_B-\eta|)\},
\end{eqnarray}
\begin{eqnarray}
M_{ef_2,N}^{LL}&=&16\sqrt{\frac{2}{3}}\pi C_Fm^4_Br_{D}\int^1_0 dx_B dzdx_3\int_0^{1/\Lambda} b_3db_3 b_Bdb_B\phi_B(x_B,b_B)\phi_D(x_3,b_3)\non
     &&\times \{[(r_{D}^2-1)(-\eta  x_3+x_3-x_B+\eta  z)]\phi_v(z) E_n(t_c)h_c(x_B,z,x_3,b_B,b_3)\non
     &&-[2 \sqrt{\eta -\eta  r_{D}^2} \phi _a(z)((\eta-1)(1-x_3) +r_{D}^2 (z-1)+x_B-z)\non
     &&-\phi _T(z) (\eta +r_{D}^2(\eta  (x_3+z-2)-x_3-x_B+1) \non
     &&-\eta (x_3+z)+x_3+x_B-1)] E_n(t_d)h_d(x_B,z,x_3,b_B,b_3)\},
\\
\non
M_{ef_2,T}^{LL}&=&16\sqrt{\frac{2}{3}}\pi C_Fm^4_Br_{D}\int^1_0 dx_B dzdx_3\int_0^{1/\Lambda} b_3db_3 b_Bdb_B\phi_B(x_B,b_B)\phi_D(x_3,b_3)\non
     &&\times \{(r_{D}^2-1)[(\eta -1) x_3+x_B+\eta  z]\phi_v(z) E_n(t_c)h_c(x_B,z,x_3,b_B,b_3)\non
     &&+[\phi _T(z) (\eta +r_{D}^2 (-x_3-x_B+1)+\eta  (r_{D}^2-1) (x_3-z)+x_3+x_B-1)\non
     &&-2 \sqrt{\eta -\eta  r_{D}^2} \phi _v(z) ((\eta-1)(1-x_3) +r_{D}^2 (z-1)+x_B-z)]\non
     &&\times E_n(t_d)h_d(x_B,z,x_3,b_B,b_3)\},
\\
\non
F_{aD,N}^{LL}&=&8\pi C_Fm^4_Bf_Br_{D}\sqrt{\eta -\eta  r_{D}^2}\int^1_0 dx_3 dz \int_0^{1/\Lambda} bdb b_3db_3\phi_D(x_3,b_3)\non
     &&\times \{[(\phi _a(z)-\phi _v(z)) (\eta -\eta  x_3+x_3)+(\phi _a(z)+\phi _v(z)) (-r_{D}^2+1)]\non
     &&\times E_a(t_e)h_e(z,x_3,b,b_3)S_t(z)+[(\phi _a(z)+\phi _v (z)) (r_{D}^2 (z-1)-z)\non
     &&+(\phi _a(z)-\phi _v (z))(\eta -1)] E_a(t_f)h_f(z,x_3,b_3,b)S_t(|\eta(x_3-1)-x_3|)\},
\\
\non
F_{aD,T}^{LL}&=&8\pi C_Fm^4_Bf_Br_{D}\sqrt{\eta -\eta  r_{D}^2}\int^1_0 dx_3 dz \int_0^{1/\Lambda} bdb b_3db_3\phi_D(x_3,b_3)\non
     &&\times \{[(\phi _a(z) -\phi _v(z))(-\eta +\eta  x_3-x_3))+(\phi _a(z)+\phi _v(z)) (-r_{D}^2+1)]\non
     &&\times E_a(t_e)h_e(z,x_3,b,b_3)S_t(z)+[(\phi _a(z)+\phi _v(z)) (r_{D}^2 (z-1)-z)+(\phi _a(z)\non
     &&-\phi _v(z)) (-\eta +1)]E_a(t_f)h_f(z,x_3,b_3,b)S_t(|\eta(x_3-1)-x_3|)\},
\\
\non
M_{aD,N}^{LL}&=&16\sqrt{\frac{2}{3}}\pi C_Fm^4_B\int^1_0 dx_B dzdx_3\int_0^{1/\Lambda} bdb b_Bdb_B\phi_B(x_B,b_B)\phi_D(x_3,b_3)\non
     &&\times \{ [2 r_{D} \phi _a(z) \sqrt{\eta -\eta  r_{D}^2}+\phi _T(z) (r_{D}^2(-\eta  (x_3+z)+x_3+\eta ^2 z-1)\non
     &&-(\eta -1) \eta  (x_B+z-1))]E_n(t_g)h_g(x_B,z,x_3,b,b_B)\non
     &&+[\phi _T(z)(-1+\eta)(-(r_{D}^2-1) r_{D}^2 x_3-\eta  ((r_{D}^2-1) z+x_B))]\non
     && \times E_n(t_h)h_h(x_B,z,x_3,b,b_B)\},
\\
\non
M_{aD,T}^{LL}&=&16\sqrt{\frac{2}{3}}\pi C_Fm^4_B\int^1_0 dx_B dzdx_3\int_0^{1/\Lambda} bdb b_Bdb_B\phi_B(x_B,b_B)\phi_D(x_3,b_3)\non
     &&\times \{ [\phi _T(z) (r_{D}^2 (-(\eta -1) (x_3+\eta  z)-1)+(\eta -1) \eta  (x_B+z-1))\non
     &&+2 r_{D} \sqrt{\eta -\eta  r_{D}^2} \phi _v(z)]E_n(t_g)h_g(x_B,z,x_3,b,b_B)+[\phi _T(z)(-1+\eta)\non
     &&\times(r_{D}^2 (x_3+\eta  z)-\eta  (z-x_B))] E_n(t_h)h_h(x_B,z,x_3,b,b_B)\}.
\end{eqnarray}
\end{enumerate}

For $D$-wave decay amplitude  $\mathcal{A}_D$, its factorization formula can be related to $\mathcal{A}_P$
by making the following replacement,
 \begin{eqnarray}
\mathcal{A}_{D}^0=\sqrt{\frac{2}{3}}\mathcal{A}_{P}^0|_{\phi_{P}^{0,s,t}\rightarrow\phi_{D}^{0,s,t} },\quad
\mathcal{A}_{D}^{\parallel,\perp}=\sqrt{\frac{1}{2}}\mathcal{A}_{P}^{\parallel,\perp}|_{\phi_{P}^{T,v,a}\rightarrow\phi_{D}^{T,v,a} }.
 \end{eqnarray}

The evolution factors $E_i(t)(i=e,a,n)$ in above equations are written as the form
\begin{eqnarray}
E_e(t)&=&\alpha_s(t) \exp[-S_B(t)-S_R(t)],\\
E_a(t)&=&\alpha_s(t) \exp[-S_D(t)-S_R(t)],\\
E_n(t)&=&\alpha_s(t) \exp[-S_B(t)-S_R(t)-S_D(t)],
\end{eqnarray}
where Sudakov exponents $S_{B,D,R}$ are defined as
\begin{eqnarray}
S_B&=& s(x_B\frac{m_B}{\sqrt2},b_B )+\frac53\int^t_{1/b_B}\frac{d\bar\mu}{\bar\mu} \gamma_q(\alpha_s(\bar\mu)),\\
S_R&=& s(z(1-r_D^2)\frac{m_B}{\sqrt2},b )+ s((1-z)(1-r_D^2)\frac{m_B}{\sqrt2},b )+ 2\int^t_{1/b}\frac{d\bar\mu}{\bar\mu} \gamma_q(\alpha_s(\bar\mu)),\\
S_D&=& s(x_3(1-\eta)\frac{m_B}{\sqrt2},b_3 ) +2\int^t_{1/b_3}\frac{d\bar\mu}{\bar\mu} \gamma_q(\alpha_s(\bar\mu)),
\end{eqnarray}
with the quark anomalous dimension $\gamma_q=-\alpha_s/\pi$ .
The explicit expressions of the functions $(s(x_B m_B/\sqrt2,b_B )\cdots)$ can be found in Appendix of Ref.~\cite{prd78-014018}.

The threshold resummation factor $S_t(x)$ is of the form:
\begin{eqnarray}
\label{eq-def-stx}
S_t(x)=\frac{2^{1+2c}\Gamma(3/2+c)}{\sqrt{\pi}\Gamma(1+c)}[x(1-x)]^c.
\end{eqnarray}
The value of $c$ is 0.3 in numerical calculations.

The hard functions $h_i$($i=a-h$) in above amplitudes can be derived from the Fourier transform of hard kernel :
\begin{eqnarray}
h_i(x1,x2(,x3),b_1,b_2)&=&h_1(\beta,b_2)\times h_2(\alpha,b_1,b_2),\\
h_1(\beta,b_2)&=&\left\{\begin{array}{ll}
K_0(\sqrt{\beta}b_2), & \quad  \quad \beta >0\\
K_0(i\sqrt{-\beta}b_2),& \quad  \quad \beta<0
\end{array} \right.\\
h_2(\alpha,b_1,b_2)&=&\left\{\begin{array}{ll}
\theta(b_2-b_1)I_0(\sqrt{\alpha}b_1)K_0(\sqrt{\alpha}b_2)+(b_1\leftrightarrow b_2), & \quad   \alpha >0\\
\theta(b_2-b_1)I_0(\sqrt{-\alpha}b_1)K_0(i\sqrt{-\alpha}b_2)+(b_1\leftrightarrow b_2),& \quad   \alpha<0
\end{array} \right.
\end{eqnarray}
where $K_0(ix)=\frac{\pi}{2}(-N_0(x)+i J_0(x))$.
$\alpha$ and $\beta$ are the factors $a_{1i}-h_{1i}$ and $e_{2i}-h_{2i}$ ($i=1,2$) given in the following paragraph.

The hard scales $t_i$ appeared in the above equations are chosen as the maximum of the virtuality of the internal momentum transition in the hard amplitudes.
For $B_{(s)}\to \bar D^{(*)}(R\to)KK$ decays, we have
\begin{eqnarray}
t_{a_1}&=&max\{m_B\sqrt{|a_{11}|},m_B\sqrt{|a_{12}|},1/b,1/b_B\},\quad t_{b_1}=max\{m_B\sqrt{|b_{11}|},m_B\sqrt{|b_{12}|},1/b,1/b_B\},\non
t_{c_1}&=&max\{m_B\sqrt{|c_{11}|},m_B\sqrt{|c_{12}|},1/b_3,1/b_B\},\quad t_{d_1}=max\{m_B\sqrt{|d_{11}|},m_B\sqrt{|d_{12}|},1/b_3,1/b_B\},\non
t_{e_1}&=&max\{m_B\sqrt{|e_{11}|},m_B\sqrt{|e_{12}|},1/b,1/b_3\},\quad t_{f_1}=max\{m_B\sqrt{|f_{11}|},m_B\sqrt{|f_{12}|},1/b,1/b_3\},\non
t_{g_1}&=&max\{m_B\sqrt{|g_{11}|},m_B\sqrt{|g_{12}|},1/b,1/b_B\},\quad t_{h_1}=max\{m_B\sqrt{|h_{11}|},m_B\sqrt{|h_{12}|},1/b,1/b_B\}, \non
\end{eqnarray}
with the factors
\begin{eqnarray}
a_{11}&=&(1-r_D^2)z, \quad\quad a_{12}=(1-r_D^2)x_Bz,\quad
b_{11}=(1-r_D^2)(x_B-\eta), \quad\quad b_{12}=a_{12},\non
c_{11}&=&a_{12}, \quad\quad c_{12}=[(1-r_D^2)z+r_D^2][x_B-(1-\eta)(1-x_3)],\non
d_{11}&=&a_{12}, \quad\quad d_{12}=(1-r_D^2)z[x_B-(1-\eta)x_3],\non
e_{11}&=&z(1-r_D^2)-1, \quad\quad e_{12}=(z-1)(r_D^2-1)[(\eta-1)x_3-\eta],\non
f_{11}&=&(1-r_D^2)[(\eta-1)x_3-\eta], \quad\quad f_{12}=e_{12},\non
g_{11}&=&e_{12}, \quad\quad g_{12}=1-[(1-z)r_D^2+z][(1-\eta)(1-x_3)-x_B],\non
h_{11}&=&e_{12}, \quad\quad h_{12}=(1-z)(1-r_D^2)[(\eta-1)x_3-\eta+x_B].
\end{eqnarray}

While for $B_{(s)}\to D^{(*)}(R\to)KK$ decays, similarly, we have
\begin{eqnarray}
t_{a_1}&=&max\{m_B\sqrt{|a_{11}|},m_B\sqrt{|a_{12}|},1/b,1/b\},\quad
t_{b_1}=max\{m_B\sqrt{|b_{11}|},m_B\sqrt{|b_{12}|},1/b,1/b_B\},\non
t_{c_1}&=&max\{m_B\sqrt{|c_{11}|},m_B\sqrt{|c_{12}|},1/b_3,1/b_B\},\quad
t_{d_1}=max\{m_B\sqrt{|d_{11}|},m_B\sqrt{|d_{12}|},1/b_3,1/b_B\},\non
t_{e_2}&=&max\{m_B\sqrt{|e_{21}|},m_B\sqrt{|e_{22}|},1/b,1/b_3\},\quad
t_{f_2}=max\{m_B\sqrt{|f_{21}|},m_B\sqrt{|f_{22}|},1/b,1/b_3\},\non
t_{g_2}&=&max\{m_B\sqrt{|g_{21}|},m_B\sqrt{|g_{22}|},1/b,1/b_B\},\quad
t_{h_2}=max\{m_B\sqrt{|h_{21}|},m_B\sqrt{|h_{22}|},1/b,1/b_B\}, \non
\end{eqnarray}
with the factors
\begin{eqnarray}
a_{11}&=&(1-r_D^2)z, \quad a_{12}=(1-r_D^2)x_Bz,\quad b_{11}=(1-r_D^2)(x_B-\eta), \quad b_{12}=a_{12},\non
c_{11}&=&a_{12}, \quad c_{12}=(1-r_D^2)z(x_B-(1-\eta)x_3),\non
d_{11}&=&a_{12}, \quad d_{12}=((z-1)r_D^2-z)[(1-\eta)(1-x_3)-x_B],\non
e_{21}&=&(1-r_D^2)[(x_3-1)\eta-x_3], \quad e_{22}=(1-\eta)(r_D^2-1)x_3z, \quad f_{21}=(\eta-1)[z+r_D^2(1-z)], \non
f_{22}&=&e_{12}, \quad g_{21}=e_{22}, \quad g_{22}=[1-(1-\eta)x_3][(1-r_D^2)z+x_B]+(1-\eta)x_3,\non
h_{21}&=&e_{22}, \quad\quad h_{22}=(1-\eta)x_3[x_B-(1-r_D^2)z].
\end{eqnarray}


\end{document}